\documentclass[prb,twocolumn,showpacs,amsmath,amssymb]{revtex4}
\usepackage[latin2]{inputenc}
\usepackage{amsmath}
\usepackage{graphicx}
\usepackage{dcolumn}
\usepackage{epstopdf}

\newcommand{\ba}{\begin{eqnarray*}}
\newcommand{\ea}{\end{eqnarray*}}
\newcommand{\baa}{\begin{eqnarray}}
\newcommand{\eaa}{\end{eqnarray}}
\def\bar{\begin{array}}
\def\ear{\end{array}}
\def\LB{\left(}
\def\RB{\right)}

\def\pr{^{\prime}}
\def\Ll{\left<}

\def\Rr{\right>}
\def\Lb{\left|}
\def\Rb{\right|}

\def\s{\sigma}
\def\f{\frac}

\begin{document}

\title{Adiabatic approximation in time-dependent reduced-density-matrix functional theory}

\author{Ryan Requist}
\email{Ryan.Requist@physik.uni-erlangen.de}
\author{Oleg Pankratov}
\affiliation{
Theoretische Festk\"orperphysik, Universit\"at Erlangen-N\"urnberg, Staudtstra\ss e 7-B2, 91058 Erlangen, Germany
}

\date{\today}

\begin{abstract}

With the aim of describing real-time electron dynamics, we introduce an adiabatic approximation for the equation of motion of the one-body reduced density matrix (one-matrix).  The eigenvalues of the one-matrix, which represent the occupation numbers of single-particle orbitals, are obtained from the constrained minimization of the instantaneous ground-state energy functional rather than from their dynamical equations.  The performance of the approximation vis-\`{a}-vis nonadiabatic effects is assessed in real-time simulations of a two-site Hubbard model.  Due to Landau-Zener-type transitions, the system evolves into a nonstationary state with persistent oscillations in the observables.  The amplitude of the oscillations displays a strongly nonmonotonic dependence on the strength of the electron-electron interaction and the rate of variation of the external potential.  We interpret an associated resonance behavior in the phase of the oscillations in terms of ``scattering'' with spectator energy levels.  To clarify the motivation for the minimization condition, we derive a sequence of energy functionals $E_v^{(n)}$, for which the corresponding sequence of minimizing one-matrices is asymptotic to the exact one-matrix in the adiabatic limit.   

\end{abstract}


\maketitle

\section{Introduction}

The ability to probe and control electronic states in molecules and nanostructures has improved considerably in recent years.\cite{brabec2000,reider2004,corkum2007,krausz2009}  These advances have highlighted the need for real-time simulations of strongly-driven and strongly-correlated electron dynamics.  Among the methods that have been used to simulate electron dynamics are the time-dependent Hartree-Fock\cite{frenkel1934,dirac1930,thouless1961} (TDHF) and multi-configuration Hartree-Fock\cite{zanghellini2004,nest2005,caillet2005} approximations, the Kadanoff-Baym equations\cite{kadanoff1962,stan2009} and Keldysh technique\cite{keldysh1964} for the nonequilibrium Green's function, and model Hamiltonian approaches.  Another family of methods, time-dependent density functional theory (TDDFT) and its extensions, describes electron dynamics in terms of single-particle densities such as the particle density and current density.  

Density functional theories enable a favorable compromise between accuracy and the computational accessibility of systems of interest by mapping the many-body Schr\"odinger equation, a high-dimensional linear problem, to the single-particle Kohn-Sham equations,\cite{kohn1965} a lower-dimensional nonlinear problem.  The Runge-Gross (RG) theorem\cite{runge1984} establishes the invertibility of the mapping --- defined via the Schr\"odinger equation --- from time-dependent local external potentials $v(\mathbf{r},t)$ to time-dependent electron densities $n(\mathbf{r},t)$, given a fixed initial state.  The invertibility of the mapping holds also in the noninteracting case, which implies that the density of an interacting system can be reproduced by an auxiliary noninteracting system with an effective potential $v_s(\mathbf{r},t)$, under mild restrictions on the initial state.\cite{vanleeuwen1999}  The noninteracting system is readily described by a set of single-particle Schr\"odinger equations, the time-dependent counterparts of the Kohn-Sham equations\cite{kohn1965} in ground-state DFT,\cite{hohenberg1964} and the density can be calculated from the resulting orbitals as $n(\mathbf{r},t) = \sum_i f_i \left| \phi_i(\mathbf{r},t)\right|^2$, where $f_i$ are occupation numbers. 

Electronic observables of the form $\hat{A} = \int d^3r a(\mathbf{r}) \hat{n}(\mathbf{r})$, e.g.~the dipole moment, are linear functionals of the density, and their time development can be obtained directly from the time-dependent density.  More general single-particle observables of the form $\hat{B} = \sum_{\s\s\pr} \int d^3r d^3r\pr b_{\s\s\pr}(\mathbf{r},\mathbf{r}\pr) \hat{\psi}_{\s}^{\dag}(\mathbf{r}) \hat{\psi}_{\s\pr}(\mathbf{r}\pr)$, where $b_{\s\s\pr}(\mathbf{r},\mathbf{r}\pr)$ is a nonlocal kernel that can contain spatial derivatives, are also functionals of the density if the external potential is local, but they are generally unknown functionals with complicated nonlinear dependence on the density.  In contrast, such observables are linear functionals of the one-body reduced density matrix (one-matrix),  
\begin{equation}
\gamma(x,x\pr;t) = \left< \Psi(t) \right| \hat{\psi}^{\dag}(x\pr) \hat{\psi}(x) \left| \Psi(t) \right>, 
\label{eqn:one-matrix}
\end{equation} 
where $x=(\mathbf{r},\s)$.  The one-matrix explicitly contains more information than the density but less information than the nonequilibrium single-particle Green's function $G^<(xt,x\pr t\pr) = i \left< \Psi \right|  \hat{\psi}^{\dag}(x\pr t\pr) \hat{\psi}(xt) \left| \Psi \right>$ ($t<t\pr$), as it is the equal-time limit of the latter, $\gamma(x,x\pr;t) = -i \lim_{t\pr\rightarrow t^+} G^<(xt,x\pr t\pr)$.

In this paper, we study correlated electron dynamics using time-dependent reduced-density-matrix functional theory,\cite{pernal2007a,appel2007,rajam2009} which is a TDDFT-like theory in which the basic variable is the one-matrix instead of the density.  The equation of motion (EOM) for the one-matrix is
\begin{align}
i &\partial_t \gamma(x_1,x_1\pr;t) 
= \big[\hat{h}_0(x_1;t)-\hat{h}_0(x_1\pr;t)\big] \gamma(x_1,x_1\pr;t) \nonumber \\ 
&+ 2\int \! dx_2 \big[v_c(x_1,x_2)-v_c(x_1\pr,x_2)\big] \Gamma(x_1,x_2,x_1\pr,x_2;t), \label{eqn:BBGKY}
\end{align}
where $\hat{h}_0(x_i,t)=\hat{\mathbf{p}}_i^2/2+v(\mathbf{r}_i,t)$, $v_c(x_1,x_2) = |\mathbf{r}_1-\mathbf{r}_2|^{-1}$, $\int dx = \sum_{\sigma} \int d^3r$ and 
\begin{align}
&\Gamma(x_1,x_2,x_1\pr,x_2\pr;t) \nonumber \\
&= \f{1}{2} \Ll \Psi(t) \Rb \hat{\psi}^{\dag}(x_2\pr) \hat{\psi}^{\dag}(x_1\pr) \hat{\psi}(x_1) \hat{\psi}(x_2) \Lb \Psi(t) \Rr
\end{align}
is the two-body reduced density matrix (two-matrix).  We use atomic units, $e^2 = \hbar = m = 1$, throughout the paper. Equation~(\ref{eqn:BBGKY}) is the first equation of the Bogoliubov-Born-Green-Kirkwood-Yvon (BBGKY) hierarchy\cite{bogoliubov1946,bogoliubov1961,bonitz1998} of equations of motion, in which the equation of motion for the $k$-body reduced density matrix contains the $(k+1)$-body reduced density matrix.  It is common to ``close'' the hierarchy at some order $k$ by expressing the $(k+1)$-body reduced density matrix in terms of the $k$-body reduced density matrix.  In principle, the hierarchy can be closed already at the first order, Eq.~(\ref{eqn:BBGKY}), because the RG theorem implies\cite{pernal2007a} that the two-matrix is a universal functional of the one-matrix, i.e., $\Gamma(t) = \Gamma([\gamma];t)$, given a fixed initial state (if a vector potential is present, the statement follows\cite{rajam2009} from the extension of RG in Ref.~\onlinecite{ghosh1988}).  This one-matrix functional approach has been applied within linear response theory to calculate frequency-dependent polarizabilities\cite{pernal2007b,pernal2007c} and excitation energies\cite{pernal2007c,giesbertz2008,giesbertz2009} of light diatomic molecules.  In the spirit of the adiabatic approximation to the linear response equations of TDDFT,\cite{zangwill1980,gross1985} where the frequency-dependent exchange-correlation (xc) kernel $f_{xc}(\omega)$ is replaced with its static limit $f_{xc}(\omega=0)$, these calculations employed only frequency-independent kernels.  For two-electron systems, an adiabatic approximation was constructed\cite{giesbertz2008,giesbertz2009} that yields the full excitation spectrum exactly, including excitations of doubly-excited character.  This represents an advantage with respect to TDDFT, where excitations of doubly-excited character are missed if a frequency-independent xc kernel is used.\cite{maitra2004,maitra2005,hirata1999,tozer2000,neugebauer2004,thiele2009b}

In real-time simulations based on Eq.~(\ref{eqn:BBGKY}), the naive adiabatic approximation consists in approximating the two-matrix functional $\Gamma([\gamma];t)$ at time $t$ by the ground-state functional $\Gamma[\gamma]$ evaluated for $\gamma = \gamma(t)$.  In other words, the time-dependent two-matrix is approximated by the \textit{adiabatic extension} of the ground-state two-matrix to the time domain.  This approximation neglects the \textit{memory dependence} of the exact functional, i.e., the exact functional $\Gamma([\gamma];t)$ will generally depend on $\gamma(t\pr)$ for all $t\pr \leq t$.  Memory effects in real-time dynamics have been studied in Refs.~\onlinecite{kurzweil2004,wijewardane2005,ullrich2006a,ullrich2006b,thiele2008,thiele2009a,khosravi2009}.  One of the motivations for taking the one-matrix as basic variable is that the universal functionals that enter the theory might have less severe memory dependence than the functionals in TDDFT (for a concrete example see Ref.~\onlinecite{rajam2009}).  Other extensions of TDDFT also hope to benefit from weaker memory dependence, notably current-density functional theory,\cite{vignale1987,vignale1988} in which the basic variable is the current density, and time-dependent deformation functional theory,\cite{tokatly2003,ullrich2006b,tokatly2007} which operates with a deformation tensor.  

The naive adiabatic extension approximation has not yet been applied in the one-matrix EOM for the following reason.\cite{appel2007,appel2008}
First, consider the eigenvalue equation $\int dx\pr \gamma(x,x\pr;t) \phi_i(x\pr,t) = f_i(t) \phi_i(x,t)$.  The eigenfunctions and eigenvalues are called natural orbitals and occupation numbers.\cite{loewdin1955}  The occupation number of a natural orbital represents its effective occupancy in the many-body wave function.  When the adiabatic extension approximation is applied to any of the available ground-state two-matrix functionals, it yields time-independent occupation numbers because the available functionals have an overly-restrictive form\cite{appel2007,appel2008} (see Sec.~\ref{ssec:IONR:AE}).  This is disappointing because although the available functionals are quite accurate for ground-state properties, their adiabatic extension misses an important aspect of the dynamics even for arbitrarily slow driving.  The exact time-dependence of the occupation numbers in model systems has been inferred by solving the many-body Schr\"odinger equation numerically.\cite{appel2008}  It was found that the occupation numbers can indeed undergo significant changes in the course of the time evolution, reflecting changes in the degree of correlation of the many-body state.

We introduce a simple modification of the adiabatic extension approximation that yields time-dependent occupation numbers even when it is applied to the available ground-state two-matrix functionals.  In this approximation, which we shall refer to as the instantaneous occupation number relaxation (IONR) approximation, the natural orbitals satisfy a time-dependent Schr\"odinger equation, while the occupation numbers are obtained ``on-the-fly'' by relaxation to the minimum of an adiabatic energy surface.  It is expected to be accurate in the adiabatic regime, i.e., when $E_{gap}\tau \gg 1$, where $E_{gap}$ is the minimum instantaneous energy gap (between the ground state and first excited state) and $\tau$ is the characteristic time scale of the external potential.  To clarify the motivation for the minimization condition and estimate the error in the resulting occupation numbers, we carry out an asymptotic analysis in the adiabatic limit $\tau \rightarrow \infty$.

We evaluate the performance of the IONR approximation by applying it to a two-site Hubbard model.  Scaling $\tau$ in various external potentials of the form $v(\mathbf{r},t/\tau)$, we find that it is fairly accurate even beyond the region of validity expected from the above adiabaticity condition.  Remarkably, it is able to describe some purely nonadiabatic\footnote{A potentially confusing term that begins with the Latin negation of a Greek negation,\textit{nonadiabatic} is used in quantum dynamics to describe corrections to purely adiabatic dynamics, i.e., corrections to the dynamics obtained by neglecting transitions between the instantaneous (adiabatic) eigenstates.}  effects, such as Landau-Zener-type (LZ) transitions.\cite{landau1932a, landau1932b, zener1932, stueckelberg1932, majorana1932}  We also assess its robustness with respect to changes in the strength of the electron-electron interaction, controlled by the Hubbard parameter $U$.  The effect of interactions on nonadiabatic dynamics is quite pronounced.  Varying $U$ across a range of values does \textit{not} result in regular, monotonic changes in the observables that exhibit nonadiabatic effects.  

The paper is organized as follows.  In Sec.~\ref{sec:IONR}, we introduce the IONR approximation and discuss its motivation and validity.  In Sec.~\ref{sec:model}, the IONR approximation and the adiabatic approximation in TDDFT are applied to a two-site Hubbard model.  We derive the effective Schr\"odinger equation for the natural orbitals (Sec.~\ref{sssec:model:IONR:Schroedinger}) and its linear and semilinear versions (Sec.~\ref{sssec:model:IONR:linearization}).  In Sec.~\ref{ssec:numerical simulations}, we carry out simulations for various time-dependent external potentials.  We compare the IONR approximation with TDHF, adiabatic TDDFT, and the numerically exact solution.  By varying $U$ and $\tau$, we study the importance of correlation for nonadiabatic effects.  Conclusions are given in Sec.~\ref{sec:conclusions}.

\section{\label{sec:IONR} Adiabatic approximations for the one-matrix equation of motion}

In this section, we discuss the adiabatic extension approximation to the one-matrix EOM and introduce a modification in which time-dependent occupation numbers are obtained ``on-the-fly'' from the constrained minimization of the ground-state energy functional.  To clarify the motivation for this minimization condition, we derive an asymptotic sequence of energy functionals in which the ground-state energy functional is the zeroth order.  The corresponding sequence of minimizing one-matrices is asymptotic to the exact one-matrix in the adiabatic limit.  Using the first-order energy functional, we estimate the error in the IONR occupation numbers.

\subsection{\label{ssec:IONR:AE} Adiabatic extension approximation}

The adiabatic extension approximation to the one-matrix EOM consists in approximating $\Gamma([\gamma];t)$ at time $t$ by the ground-state reconstruction $\Gamma[\gamma]$ evaluated for $\gamma = \gamma(t)$, the self-consistently evolving one-matrix at the same time.  The existence of $\Gamma[\gamma]$, which is called a \textit{reconstruction} of the two-matrix, is implied by Gilbert's extension\cite{gilbert1975} of the Hohenberg-Kohn theorem.\cite{hohenberg1964}  Gilbert proved that the ground-state wave function is a universal functional of the one-matrix.  Approximate reconstructions are available from many of the existing ground-state energy functionals in reduced-density-matrix functional theory (RDMFT), where they are used to derive approximations to the electron-electron interaction energy functional $W[\gamma] = \mathrm{Tr}(\hat{W}\hat{\Gamma})$ ($\hat{W}$ represents the Coulomb interaction operator).  In RDMFT, the analog of the Hohenberg-Kohn energy functional is $E_v[\gamma] = T[\gamma] + V[\gamma] + W[\gamma]$, where $T[\gamma]$ is the kinetic energy and $V[\gamma] = \mathrm{Tr}(\hat{v} \hat{\gamma})$ is the external potential energy.  

All of the available ground-state reconstructions have an overly restrictive form that couples only to the \textit{two-index} Coulomb integrals.  These include the direct $U_{ijij}$ and exchange $U_{ijji}$ Coulomb integrals, as well as a third type, $U_{iijj}$, where $U_{ijkl} \equiv \left< \phi_i \phi_j \left| \hat{v}_c \right| \phi_k \phi_l\right>$ (in this definition the orbitals do not contain spin factors).  
Reconstructions that couple only to such Coulomb integrals will be called \textit{two-index reconstructions}.  It will be convenient to express the two-matrix in the natural orbital basis according to
\begin{align}
\Gamma_{ijkl}(t) &= \int dx_1 dx_2 dx_1\pr dx_2\pr \;\phi_i^*(x_1,t) \phi_j^*(x_2,t)  \nonumber \\
& \times \Gamma(x_1,x_2,x_1\pr,x_2\pr;t) \phi_k(x_1\pr,t) \phi_l(x_2\pr,t). 
\end{align}
If the wave function is an eigenstate of $\hat{S}_z$, all of the natural orbitals have a pure spin state, spin-up or spin-down.  For two-index reconstructions, the most general forms for the spin-parallel ($\sigma \sigma$) and spin-anti-parallel ($\sigma \overline{\sigma}$) elements are\cite{kollmar2004,piris2006}
\begin{align}
\Gamma_{ijkl}^{\sigma \sigma \sigma \sigma} &= F_{ij}^{\sigma \sigma} \delta_{ik} \delta_{jl}  + G_{ij}^{\sigma \sigma} \delta_{il} \delta_{jk} \nonumber \\ 
\Gamma_{ijkl}^{\sigma \overline{\sigma} \sigma \overline{\sigma}} &= F_{ij}^{\sigma \overline{\sigma}} \delta_{ik} \delta_{jl} + G_{ij}^{\sigma \overline{\sigma}} \delta_{il} \delta_{jk} + 
H_{ik}^{\sigma \overline{\sigma}} \delta_{ij} \delta_{kl},\label{eqn:reconstruction}
\end{align}
where the spin indices are made explicit.  Note that the $\Gamma_{ijkl}^{\sigma \overline{\sigma} \,\overline{\sigma} \sigma}$ elements are not independent because $\Gamma_{ijkl}^{\sigma \overline{\sigma} \,\overline{\sigma} \sigma} = -\Gamma_{ijlk}^{\sigma \overline{\sigma} \sigma \overline{\sigma}}$.  For spin-unpolarized systems, the symmetry properties of the two-matrix require all $F$'s and $G$'s to be real symmetric and $H_{ij}^{\sigma \overline{\sigma}}$ to be Hermitian.  In all of the available reconstructions, the $F$'s, $G$'s and $H_{ij}^{\sigma \overline{\sigma}}$ are taken to be real-valued functions of the occupation numbers.\footnote{When the Hamiltonian is real and its ground state is nondegenerate, the exact reconstruction must be real.}  For example, in the M\"uller functional $F_{ij}^{\sigma\sigma\pr} = f_{i\sigma} f_{j\sigma\pr}$, $G_{ij}^{\sigma\sigma\pr} = -\sqrt{f_{i\sigma} f_{j\sigma\pr}} \delta_{\sigma \sigma\pr}$, $H_{ij}^{\sigma\overline{\sigma}}=0$.  Using the adiabatic extension of Eq.~(\ref{eqn:reconstruction}) with real-valued $F$'s and $G$'s in the one-matrix EOM, we obtain [see Eq.~(\ref{eqn:occnum:evolution}) below]
\begin{align}
\partial_t f_{k\sigma} &= 4 \;\mathrm{Im}\;\sum_{\s\pr} \sum_{ijl} \Gamma_{ijkl}^{\s \s\pr \s \s\pr} U_{klij} \nonumber \\
&= 4 \; \mathrm{Im} \; \sum_i H_{ik}^{\s \overline{\s}} \;  U_{kkii}. \label{eqn:occnum:two:index}
\end{align}
Most of the available reconstructions have $H_{ij}^{\s \overline{\s}}=0$ as in the M\"uller functional.  Such reconstructions will be called \textit{JK-only}\cite{cioslowski2003} reconstructions because they couple only to the direct and exchange Coulomb integrals.  For such reconstructions the right-hand side of Eq.~(\ref{eqn:occnum:two:index}) vanishes identically,\cite{pernal2007b,appel2007,appel2008} so that the occupation numbers remain constant in time for \textit{any} external potential $v(\mathbf{r},t)$, regardless of its strength and rate of change.  This is clearly a bad approximation, as even in the adiabatic regime the occupation numbers can undergo large net changes.  

\subsection{\label{ssec:dyneqns} Dynamical equations for the natural orbitals and occupation numbers}

Expressing the one-matrix EOM, Eq.~(\ref{eqn:BBGKY}), in terms of the natural orbitals and occupation numbers, leads to the following dynamical equations:\cite{pernal2007a,appel2007}
\begin{align}
i \partial_t \left| \phi_k \right> =  \left( \hat{t} + \hat{v}_{\mathrm{eff}} \right) \left| \phi_k \right>, \label{eqn:orb:evolution}
\end{align}
\begin{align}
i \partial_t f_k &= 4 \;\mathrm{Im} \sum_{ijl} \Gamma_{ijkl}(t) \left< \phi_k \phi_l \left| \hat{v}_c \right| \phi_i \phi_j\right>  \nonumber \\
&= \left< \phi_k \left| \hat{u} \right| \phi_k \right>, \label{eqn:occnum:evolution}
\end{align}
where $\hat{t} = -\nabla^2/2$ is the kinetic energy operator and $\hat{v}_{\mathrm{eff}}$ and $\hat{u}$ are integral operators with kernels
\begin{align}
v_{\mathrm{eff}}(x_1,x_1\pr) &= v(r_1,t) \delta(r_1-r_1\pr) \nonumber \\
&+ \sum_{jk}\pr \f{u_{jk}}{f_k-f_j} \phi_j(x_1,t) \phi_k^*(x_1\pr,t), \nonumber \\
u(x_1,x_1\pr) &= 2\int dx_2 \left[ v_c(x_1,x_2)-v_c(x_1\pr,x_2) \right]\nonumber \\
&\times \Gamma(x_1,x_2,x_1\pr,x_2;t), \label{eqn:veff}
\end{align}
where $u_{jk} = \Ll \phi_j \Lb \hat{u} \Rb \phi_k \Rr$ and the prime indicates that the sum over $k$ is restricted to $k$ for which $f_k \neq f_j$.  As in the static case,\cite{pernal2005} the components $v_{\mathrm{eff},jk}$ for any $j$ and $k$ for which $f_j = f_k$, as well as the diagonal components $v_{\mathrm{eff},kk}$, are not uniquely defined; however, this is of no consequence as these components have no effect on the dynamics of the one-matrix.  Changing the diagonal components $v_{\mathrm{eff},kk}$ leads to a redefinition of the overall time dependent phase of $\phi_k$, which cancels out in the expression $\gamma(x,x\pr;t) = \sum_k f_k(t) \phi_k(x,t) \phi_k^*(x\pr,t)$.  Changing the former components simply mixes natural orbitals $\phi_j$ and $\phi_k$ for which $f_j = f_k$, again leaving the one-matrix invariant.  Such occupationally-degenerate natural orbitals are defined, in the first place, only up to an arbitrary unitary transformation within the degenerate subspace.  Symmetry is a common source of occupational degeneracy.\cite{kutzelnigg1963}  The natural orbitals satisfy an effective Schr\"odinger equation, while the occupation numbers satisfy a dynamical equation in which the kinetic energy operator and external potential do not appear.  Equations~(\ref{eqn:orb:evolution}-\ref{eqn:veff}) reveal a partitioning of the operator $\hat{u}$:  only its off-diagonal part, which is Hermitian, contributes in Eq.~(\ref{eqn:orb:evolution}), while only its diagonal part, which is purely imaginary, contributes in Eq.~(\ref{eqn:occnum:evolution}).

Equation (\ref{eqn:orb:evolution}) can be interpreted as the single-particle Schr\"odinger equation for the natural orbitals of an auxiliary noninteracting system (a generalized KS system) that reproduces the time-dependent one-matrix of the interacting system.  As in the ground-state theory,\cite{gilbert1975} the effective potential $\hat{v}_{\textrm{eff}}$ is nonlocal even though the given external potential is local.  The occupation numbers are generally fractional ($0\leq f_i \leq 1$).  Therefore, the generalized KS system should be interpreted as an ensemble state.\cite{requist2008}  The occupation numbers are related to the ensemble weights, and their time-dependence can be attributed to the coupling to a fictitious environment through the nonhermitian diagonal elements of $\hat{u}$. 

\subsection{\label{ssec:IONR:IONR} IONR approximation}

We propose a modification of the adiabatic extension approximation that generates time-dependent occupation numbers even for JK-only reconstructions.  The method consists in propagating Eq.~(\ref{eqn:orb:evolution}) in the adiabatic extension approximation, while at each instant evaluating $\hat{v}_{\rm eff}$ at the occupation numbers that minimize the ground-state energy functional $E_v[\gamma]$ subject to the constraint that the natural orbitals are equal to the time-dependent natural orbitals at that same instant.  It can be interpreted as an adiabatic extension with respect to the orbital degrees of freedom coupled to a condition of instantaneous relaxation to the minimum of an adiabatic effective energy surface for the occupation numbers.  Fundamentally, the reason that the dynamical equation for the occupation numbers is replaced by a minimum condition, which is not a differential equation in the time variable, is that the adiabatic limit is a singular limit of the time-dependent Schr\"odinger equation.  In this paper, we formulate the approximation for systems that start in the ground state at the initial time.  Therefore, the occupation numbers relax on an effective energy surface defined by the ground-state energy functional.  One could also consider the dynamics of a system that starts in another adiabatic eigenstate and determine the occupation numbers from relaxation to a minimum or stationary point on the corresponding adiabatic energy surface, provided such a surface exists locally and obeys a local minimum condition or stationary condition.  This is true, for instance, for the lowest energy state of a given symmetry.\cite{gunnarsson1976}  An important feature of the IONR approximation is that the condition of instantaneous relaxation introduces a strong temporal coherence between the orbitals and occupation numbers.  

The IONR approximation is conceptually similar to an adiabatic approximation within linear-response time-dependent reduced-density-matrix functional theory developed and tested in Refs.~\onlinecite{pernal2007b,pernal2007c,giesbertz2008,giesbertz2009}.  It was found that the occupation numbers had vanishing linear response in the adiabatic extension approximation.  Therefore, the \textit{static} linear response equation for the occupation numbers, which gives nonzero response, was extended to finite $\omega$ and incorporated into the frequency-dependent linear response equations for the natural orbitals.  This assumes that the occupation numbers respond instantaneously to the time-dependent perturbation and is equivalent to the result that would be obtained from the linear response of the IONR approximation.  

\subsection{\label{ssec:IONR:justification} Minimum principles and the motivation for the IONR approximation}

The IONR approximation can be motivated by an asymptotic analysis of the many-body Schr\"odinger equation in the adiabatic limit $\tau \rightarrow \infty$.  On the basis of such an analysis, we identify a sequence of ``ground-state'' energy functionals $E_v^{(n)}[\gamma]$, each of which satisfies a local minimum principle at each instant of time for sufficiently large $\tau$.  The instantaneous minimization of $E_v^{(n)}[\gamma]$ gives an approximation $\gamma^{(n)}$ with error of order $\tau^{-(n+1)}$ to the exact time-dependent one-matrix.  In the IONR approximation, the occupation numbers are calculated from the constrained minimization of the Hohenberg-Kohn-like functional $E_v[\gamma]$, which is the zeroth-order member of the sequence.  An explicit comparison of the zeroth-order and first-order energy functionals affords us a means of estimating the error in the IONR occupation numbers.

Consider a system with Hamiltonian of the form $\hat{H}=\hat{H}(t/\tau)$ that starts in the ground state at $t=-\infty$.  Further, suppose that $\hat{H}(t/\tau)$ is infinitely differentiable with respect to $t$ and that the instantaneous ground state remains gapped for all time.  Following Ref.~\onlinecite{berry1987}, we perform successive unitary transformations $\hat{U}^{(n)} = \hat{U}^{(n)}(t)$, each attempting to approach with increasing accuracy the exact time evolution operator $\hat{U}=\hat{U}(t)$; $\hat{U}(-\infty) = 1$.  Each transformation is determined from quasistatic considerations.  To define the zeroth-order transformation, we first require that it diagonalizes the Hamiltonian at each instant of time, i.e., $\hat{U}^{(0)\dag} \hat{H} \hat{U}^{(0)} = \hat{E}^{(0)}$.  This determines $\hat{U}^{(0)}$ up to time-dependent phases that we fix as follows.  Note that $\hat{U}^{(0)}$ propagates the instantaneous (zeroth-order adiabatic) eigenstates $\big|\psi_i^{(0)}\big>$ of $H$, i.e., $\big|\psi_i^{(0)}(t)\big> = \hat{U}^{(0)}(t) \big| \psi_i^{(0)}(-\infty)\big>$.  Hence, the additional requirement $\mathrm{Im} \big< \psi_i^{(0)} \big| \partial_t \psi_i^{(0)}\big> = 0$ for all $i$, which corresponds to parallel transport,\cite{berry1987} determines $U^{(0)}$ uniquely.  Substituting $\big|\psi\big> = \hat{U}^{(0)} \big|\psi^{(1)}\big>$ in $i \partial_t \big|\psi\big> = \hat{H} \big|\psi\big>$, we obtain
\begin{align}
&i \partial_t \big| \psi^{(1)} \big> = \hat{H}^{(1)} \big| \psi^{(1)} \big>; \nonumber \\ 
&\hat{H}^{(1)} = \hat{E}^{(0)} - i \tau^{-1} \hat{U}^{(0)\dag} \partial_s \hat{U}^{(0)}, \label{eqn:Schroed:second} 
\end{align}
where $s=t/\tau$.  The term $\hat{A}^{(1)} \equiv -i \hat{U}^{(0)\dag} \partial_s \hat{U}^{(0)}$, which is called the \textit{nonadiabatic coupling}, is responsible for transitions between adiabatic eigenstates.  In the basis of zeroth-order adiabatic states, it has the elements\footnote{In the Born-Oppenheimer approximation, the adiabatic electronic eigenstates $\Lb \psi_i \Rr$ depend implicitly on the nuclear trajectories $\mathbf{R}_{n}(t)$, so that $\tau^{-1} A_{ij}$ contains the contribution $\sum_n  \Ll \psi_i \big| \mathbf{V}_{n} \cdot \mathbf{P}_{n} \big| \psi_j \Rr$, where $\mathbf{V}_{n} = d\mathbf{R}_n/dt$ and $\mathbf{P}_{n} = -i \nabla_{n}$.  In the literature, the factor $\Ll \psi_i \big| \nabla_{n} \psi_j \Rr$ is often referred to as the \textit{nonadiabatic coupling vector}.}
\begin{align}
A_{ij} &= -i \big<\psi_i \big| \partial_s \psi_j \big> \nonumber \\
&= i \f{\big< \psi_i \big| \partial_s\hat{H} \big| \psi_j \big>}{E_i-E_j}, \qquad (j\neq i) \label{eqn:A}
\end{align}
where the superscripts have been omitted.  Due to the parallel transport condition, $\hat{A}$ is purely off-diagonal.  We now consider the adiabatic energies of $\hat{H}^{(1)}$.  Expanding the lowest energy eigenvalue of $\hat{H}^{(1)}$ with respect to $\tau^{-1}$, we find
\begin{align}
E_0^{(1)} = E_0^{(0)} + \tau^{-2} \sum_{i\neq 0} \frac{A_{0i}^{(1)} A_{i0}^{(1)}}{E_0^{(0)} - E_i^{(0)}} + \mathcal{O}(\tau^{-3}).\label{eqn:energy:1}
\end{align}
The second term is an $\mathcal{O}(\tau^{-2})$ shift of the instantaneous ground-state energy induced by the nonadiabatic coupling.  Apart from an overall sign, it is identical to the induced inertia term that enters in the effective Schr\"odinger equation for the nuclear wave function in the Born-Oppenheimer approximation.\cite{goldhaber2005}  The first-order adiabatic eigenstate corresponding to $E_0^{(1)}$ is
\begin{align}
\psi_0^{(1)} = \psi_0^{(0)} + \tau^{-1} \sum_{i\neq 0} \f{A_{i0}^{(1)}}{E_0^{(0)} - E_i^{(0)}} \psi_i^{(0)} + \mathcal{O}(\tau^{-2}) {.} \label{eqn:psi1}
\end{align} 
In the absence of magnetic fields, $\hat{H}$ is real and all of the zeroth-order eigenstates $\psi_i^{(0)}$ can be chosen to be real.  Therefore, $\psi_0^{(1)}$ has the same density, to order $\tau^{-1}$, as the state $\psi_0^{(0)}$ because the nonadiabatic coupling $\hat{A}^{(1)}$ is purely imaginary.  Moreover, since $\psi_0^{(0)}$ is assumed to be nondegenerate for all time, it must have everywhere vanishing current.  The terms containing $\hat{A}^{(1)}$ in Eq.~(\ref{eqn:psi1}) generate a current of order $\tau^{-1}$.  If $\psi_0^{(1)}$ were used to construct the two-matrix, Eq.~(\ref{eqn:occnum:evolution}) would generate time-dependent occupation numbers.  In principle, developing the asymptotic series for the wave function in powers of $\tau^{-1}$ gives a way to derive systematic corrections to the ground-state reconstruction.  

Iteration\cite{berry1987} of the above diagonalization procedure together with the definitions $\big|\psi^{(n)}\big> = \hat{U}^{(n)} \big|\psi^{(n+1)}\big>$ and $\mathrm{Im} \big< \psi_i^{(n)} \big| \partial_t \psi_i^{(n)}\big> = 0$ gives the $n$th-order Hamiltonian $\hat{H}^{(n)} = \hat{E}^{(n-1)} - i \tau^{-1} \hat{U}^{(n-1)\dag} \partial_s \hat{U}^{(n-1)}$.  
The sequence of unitary transformations can be understood as an attempt to transform the Schr\"odinger equation to a basis in which the nonadiabatic coupling is as small as possible.  The $n$th-order adiabatic state $\big|\psi_0^{(n)}\big>$ is an approximation for the wave function (up to an overall phase) with error of order $\tau^{-(n+1)}$ for all time.  However, as described in Ref.~\onlinecite{berry1987}, the sequence of approximants $\big|\psi_0^{(n)}\big>$ cannot converge uniformly to the exact solution (it ultimately diverges as $n!$), for if it did there could be no nonadiabatic (Landau-Zener-type) transitions since every member of the sequence is asymptotic to the same zeroth-order adiabatic eigenstate at $t = \infty$ as at $t = -\infty$.\cite{berry1987}  The approximant $\big|\psi_0^{(n)}\big>$ for which the error is minimum may provide an accurate approximation for the wave function, but it does not describe nonadiabatic transitions.  Assuming that the time dependence of $\hat{H}$ is sufficiently smooth, nonadiabatic transitions are nonperturbative, i.e., beyond any power of $\tau^{-1}$.  

Let us consider the IONR approximation from the perspective of the above asymptotic analysis.  The lowest eigenvalues of the sequence $\hat{H}^{(n)}$ provide a sequence of adiabatic ``ground-state'' energies $E_0^{(n)}=E_0^{(n)}(t)$.  The zeroth-order energy $E_0^{(0)}=E_0^{(0)}(t)$ is simply the instantaneous minimum of $E_v[\gamma]$ for $v=v(t)$.   In a forthcoming article,\footnote{R. Requist and O. Pankratov, \textit{in preparation}.} we shall show that under certain conditions there is a sequence of energy functionals $E_v^{(n)}[\gamma]$ corresponding to the $E_0^{(n)}$.  In contrast to the HK-like energy functional $E_v[\gamma]$, for which we can subtract away the contribution of the external potential energy $V[\gamma]$, leaving a \textit{universal} functional $F[\gamma] = E_v[\gamma] - V[\gamma]$ that does not depend on $v$, the higher-order energy functionals have a nonlinear dependence on $v$ and its derivatives with respect to time, so that the the contribution of the external potential cannot be separated.  For large enough $\tau$, each $E_v^{(n)}[\gamma]$ satisfies a local minimum principle yielding $E_0^{(n)}$ for $\gamma = \gamma^{(n)}$, where $\gamma^{(n)}$ is the one-matrix corresponding to $\big|\psi_0^{(n)}\big>$.  The $\gamma^{(n)}$ are approximations with error of order $\tau^{-(n+1)}$ to the exact time-dependent one-matrix.  The $E_v^{(n)}[\gamma]$ can be thought of as defining adiabatic energy surfaces.  In principle, the constrained minimization of the IONR approximation could be performed on one of these higher-order energy surfaces.  Since nonadiabatic transitions are not captured in the above asymptotic analysis, we might expect them to be missing in the IONR approximation as well.  However, the effects of nonadiabatic transitions are partially accounted for through the effective Schr\"odinger equation for the natural orbitals, and the results that we shall present in Sec.~\ref{ssec:numerical simulations} suggest that nonadiabatic transitions are, in fact, fairly well represented in the IONR approximation.

\subsection{\label{ssec:IONR:error} Error estimate for the IONR occupation numbers}

We can use the first-order energy surface to estimate the error in the IONR occupation numbers.  Let the one-matrix obtained in the IONR approximation be denoted $\tilde{\gamma}$.  To estimate the error in the occupation numbers $\tilde{f}_i$, we take $\gamma^{(1)}$ as a reference and calculate the linear deviations with respect to it.  Since the error in $\gamma^{(1)}$ is $\mathcal{O}(\tau^{-2})$, the absolute error of the $\tilde{f}_i$ is $\big|\tilde{f}_i-f_i^{(1)}\big| + \mathcal{O}(\tau^{-2})$.  We define the functional $G_v^{(1)}[\gamma] = E_v^{(1)}[\gamma] - \mu^{(1)} \int dx [\gamma(x,x)-N]$, where $\mu^{(1)}$ is a Lagrange multiplier that maintains the total particle number $N$.  We assume $G_v^{(1)}[\gamma]$, at its minimum, satisfies the stationary condition $\delta G_v^{(1)} = 0$ for an arbitrary variation $\delta \gamma$, apart from variations in occupation numbers that are exactly $0$ or $1$.  Natural orbitals with occupation numbers exactly $0$ or $1$ are called pinned states, and the energy need not be stationary with respect to the variations of pinned occupation numbers.\cite{lathiotakis2007,helbig2007,requist2008}  Under the above assumption, we have the stationary condition (for all $t$) 
\begin{align}
\left.\frac{\partial G_v^{(1)}}{\partial f_k}\right|_{\gamma = \gamma^{(1)}(t)}  = 0 \label{eqn:min:1}
\end{align}
for all unpinned occupation numbers.  Similarly, in the IONR approximation, we have the stationary condition
\begin{align}
\left.\frac{\partial G_v}{\partial f_i}\right|_{\gamma = \tilde{\gamma}(t)} = 0,
\end{align}
where $G_v = E_v - \mu \int dx [\gamma(x,x)-N]$.  Shifting the evaluation point in Eq.~(\ref{eqn:min:1}) from $\gamma^{(1)}(t)$ to $\tilde{\gamma}(t)$, we obtain to lowest order
\begin{align}
0 &= \tau^{-2} \frac{\partial \Omega^{(2)}}{\partial f_k} \!+ \sum_i \frac{\partial^2 E_v^{(1)}}{\partial f_k\partial f_i} \left[f_i^{(1)}(t) - \tilde{f}_i(t)\right] \nonumber \\
&+ \sum_i \int dx \frac{\partial}{\partial f_k} \frac{\delta E_v^{(1)}}{\delta \phi_i(x)} 
\left[\phi_i^{(1)}(x,t) - \tilde{\phi}_i(x,t)\right] 
\nonumber \\
&+ \sum_i \int dx \frac{\partial}{\partial f_k} \frac{\delta E_v^{(1)}}{\delta \phi_i^*(x)} \left[\phi^{*(1)}_i(x,t) - \tilde{\phi}_i^*(x,t)\right], \label{eqn:min:shifted}
\end{align}
where $\Omega^{(2)} = \tau^{2} (G_v^{(1)} - G_v)$ and all derivatives are evaluated for $\gamma = \tilde{\gamma}$.  Furthermore, in all second derivative terms we have been able to replace $G_v^{(1)}$ by $E_v^{(1)}$ because the shift $\Delta \gamma = \tilde{\gamma} - \gamma^{(1)}$ is number conserving.  Defining $\Delta f_i = \tilde{f}_i - f_i^{(1)}$, $\Delta \phi_i = \tilde{\phi}_i - \phi_i^{(1)}$ and introducing the anti-Hermitian matrix $\Delta c_{ji} = \big< \tilde{\phi}_j \big| \Delta \phi_i \big>$, Eq.~(\ref{eqn:min:shifted}) becomes
\begin{align}
-\tau^{-2} \left.\frac{\partial \Omega^{(2)}}{\partial f_k}\right|_{\gamma =\tilde{\gamma}(t)} &\!\!\!= \sum_i \chi_{kkii}^{-1} \; \Delta f_i \nonumber \\
&\!\!\!+ \sum_{i,j\neq i} \left[ \chi_{kkji}^{-1} \; f_i \: \Delta c_{ji} +\chi_{kkij}^{-1} \; f_i \: \Delta c_{ji}^* \right] {.} \label{eqn:min:shifted:2}
\end{align}
We have used 
\begin{align}
 \left. \frac{\delta^2 E_v^{(1)}}{\delta \gamma(x_1\pr x_1) \delta \gamma(x_2\pr x_2)}\right|_{\gamma = \tilde{\gamma}} &\approx 
 \left. \frac{\delta^2 E_v}{\delta \gamma(x_1\pr x_1) \delta \gamma(x_2\pr x_2)}\right|_{\gamma = \gamma^{(0)}} \nonumber \\
&= -\chi^{-1}(x_1 x_1\pr x_2 x_2\pr) {,}
\end{align}
where $\chi^{-1}$ is the inverse static response function of the instantaneous ground state.  Here, $E_v^{(1)}$ has been replaced by $E_v$ and the evaluation point has been changed from $\tilde{\gamma}$ to $\gamma^{(0)}$, which both introduce higher-order errors.  In the basis of natural orbitals, the response function $\chi(x_1 x_1\pr, x_2 x_2\pr) = \delta \gamma(x_1 x_1\pr)/\delta v(x_2 x_2\pr)$ is expressed as
\begin{align}
\chi_{ijkl} &= \int dx_1 dx_1\pr dx_2 dx_2\pr \; \phi_i^*(x_1) \phi_j(x_1\pr) \nonumber \\ 
&\times \; \chi(x_1 x_1\pr, x_2 x_2\pr) \;\phi_k(x_2\pr) \phi_l^*(x_2).
\end{align}
Equation (\ref{eqn:min:shifted:2}) is a set of linear equations relating the linear deviations $\Delta f_i$ to $\Delta c_{ij}$.  It is coupled to another set of linear equations obtained from an analysis of the linear deviations in the effective Schr\"odinger equation.  Together these constitute a linear system of equations that can be solved for the $\Delta f_i$ and $\Delta \phi_i$. 

An important simplification can be obtained by considering the case that the electron-electron interaction is weak.  Suppose a coupling constant $U$ is introduced into the Coulomb interaction.  Consider the following four subblocks of $\chi$: 1) subblock AA, $\chi_{iijj}$, 2) subblock AB, $\chi_{iikl}$ ($k\neq l$), 3) subblock BA and $\chi_{klii}$ ($k\neq l$), and 3) subblock BB, $\chi_{ijkl}$ ($i\neq j$, $k\neq l$).  If the system remains gapped in the limit $U \rightarrow 0$, subblock AA is $\mathcal{O}(U^{2})$, subblocks AB and BA are $\mathcal{O}(U^{1})$, and subblock BB is $\mathcal{O}(1)$.  Therefore, $\chi_{iijj}^{-1} = \mathcal{O}(U^{-2})$, while $\chi_{iikl}^{-1}$ and $\chi_{klii}^{-1}$ are $\mathcal{O}(U^{-1})$ and $\chi_{ijkl}^{-1}$ is $\mathcal{O}(1)$.  When only the former are retained, Eq.~(\ref{eqn:min:shifted:2}) gives the following estimate for the error of the IONR occupation numbers:
\begin{align}
\Delta f_i &= - \tau^{-2} \sum_k X_{ik} \left.\frac{\partial (E_v^{(1)} - E_v)}{\partial f_k}\right|_{\gamma =\gamma^{(0)}} \nonumber \\
&= \mathcal{O}(U^2/\tau^2),\label{eqn:error}
\end{align}
where $X = (\overline{\chi^{-1}})^{-1}$ and $(\overline{\chi^{-1}})_{ijkl} = \delta_{ij} \delta_{kl} \chi_{ijkl}^{-1}$.  The functional $E_v^{(1)} - E_v$ is a ``warping'' of the zeroth-order energy surface, and $X$ can be interpreted as an effective response function for the occupation number degrees of freedom.  Since the error of $\gamma^{(1)}$ is also $\mathcal{O}(U^2/\tau^2)$, the absolute error in the IONR occupation numbers is $\mathcal{O}(U^2/\tau^2)$.  For the model system that we consider in Sec. \ref{sec:model}, we have verified numerically that the error is indeed $\mathcal{O}(U^2/\tau^2)$.

\section{\label{sec:model} Application to a model system}

In this section, we apply the IONR approximation and the adiabatic extension approximation in TDDFT (ADFT) to a simple model system.  By varying the Hubbard parameter $U$ and the characteristic time scale $\tau$ of the external potential, we can study the interplay of electron interactions and nonadiabatic dynamics.  

We consider the two-site Hubbard model with two electrons.\footnote{Electron dynamics in small Hubbard clusters was studied within many-body perturbation theory in Refs.~\onlinecite{verdozzi2008} and \onlinecite{puigvonfriesen2009}.  The static version of the two-site Hubbard model was studied within reduced-density-matrix functional theory in Ref.~\onlinecite{requist2008}.}  The Hamiltonian is
\begin{align}
\hat{H} = \f{1}{2} V_1 \sum_{\s} \big( \hat{c}_{1\s}^{\dag} \hat{c}_{2\s} + \hat{c}_{2\s}^{\dag} \hat{c}_{1\s} \big) + \f{1}{2} V_3 \big( \hat{n}_1 - \hat{n}_2 \big) + \hat{U} 
\end{align}
where $\hat{c}_{i\s}^{\dag}$ and $\hat{c}_{i\s}$ are the creation and annihilation operators of an electron with spin $\s$ in site $i$, $\hat{U} =  U (\hat{n}_{1\uparrow} \hat{n}_{1\downarrow} + \hat{n}_{2\uparrow} \hat{n}_{2\downarrow})$, and we have used $V_1/2$ instead of the usual notation $-t$ for the hopping parameter.  In this model, the analog of the local external potential $v(\mathbf{r},t)$ is a time-dependent bias $V_3=V_3(t/\tau)$.  It will be convenient to write the Hamiltonian as 
\begin{align}
\hat{H} = \f{1}{2} \vec{V} \cdot \hat{\vec{\s}} + \hat{U}, \label{eqn:H:Pauli}
\end{align}
where $\vec{V} = (V_1, 0, V_3)$ and $\hat{\vec{\sigma}} = (\hat{\s}_1, \hat{\s}_2, \hat{\s}_3)$ is the second-quantization representation of the Pauli matrices, i.e., $\hat{\sigma}_i = \sum_{\s} (\hat{c}_{1\s}^{\dag},\: \hat{c}_{2\s}^{\dag}) \sigma_i \LB \bar{c} \hat{c}_{1\s} \\ \hat{c}_{2\s} \ear \RB$.  The operator $\hat{\vec{\sigma}}$ is not related to physical spin; below, it will be identified with a Bloch pseudospin.  In this paper, we consider only constant $V_1$ and $V_2=0$. Generalizing Eq.~(\ref{eqn:H:Pauli}) by letting $V_1$ and $V_2$ be time-dependent functions is roughly analogous to introducing a time-dependent vector potential in the case of continuous variables.  

We assume that the initial state is a spin-singlet state with $S_z=0$.  As the external potential is spin-independent and the singlet and triplet sectors are decoupled, this spin configuration will be preserved for all times.  In the basis $\big\{\hat{c}_{1\uparrow}^{\dag} \hat{c}_{1\downarrow}^{\dag} \Lb 0\Rr, \;\f{1}{\sqrt{2}}(\hat{c}_{1\uparrow}^{\dag} \hat{c}_{2\downarrow}^{\dag} + \hat{c}_{2\uparrow}^{\dag} \hat{c}_{1\downarrow}^{\dag}) \Lb 0 \Rr, \; \hat{c}_{2\uparrow}^{\dag} \hat{c}_{2\downarrow}^{\dag} \Lb 0\Rr\big\}$, the Hamiltonian is
\baa
H = \LB \bar{ccc} 
U + V_3 & \f{1}{\sqrt{2}}(V_1 - i V_2) & 0 \\  
\f{1}{\sqrt{2}}(V_1 + i V_2) & 0 & \f{1}{\sqrt{2}}(V_1 - i V_2) \\
0 & \f{1}{\sqrt{2}}(V_1 + i V_2) & U - V_3 \ear \RB. \nonumber
\eaa
The instantaneous eigenenergies, which are also referred to as \textit{adiabatic energy levels}, are the following roots of the cubic secular equation:
\begin{align}
E_1 &= \f{2}{3} \left[U - \rho \; \cos\LB \f{\omega}{3} \RB \right]\nonumber \\
E_2 &= \f{2}{3} \left[U - \rho \; \cos\LB \f{\omega}{3}-\f{2\pi}{3} \RB \right] \nonumber \\
E_3 &= \f{2}{3} \left[U - \rho \; \cos\LB \f{\omega}{3}-\f{4\pi}{3} \RB \right], 
\label{eqn:energies}
\end{align}
where
\begin{align}
\rho^2 &= -9 Q = U^2 + 3 \vec{V}^2 \nonumber \\
\kappa^3 &= -54 R = U \left[ 2U^2 + 9 (\vec{V}^2 - 3 V_3^2)\right]\label{eqn:rho}
\end{align}
and 
\begin{equation}
\omega = \cos^{-1}\LB\f{R}{\sqrt{-Q^3}}\RB = \cos^{-1}\LB\f{\kappa^3}{2 \rho^3}\RB. 
\label{eqn:phi}
\end{equation}
The definitions of the variables $Q$ and $R$ are conventional for cubic equations.  The instantaneous eigenstates $\Lb \psi_i \Rr$ can be expressed as 
\begin{equation}
\Lb \psi_i \Rr = N_i \LB \bar{c}
- \left[|W|^2 + E_i (U - V_3 - E_i)\right] e^{-i \varphi_0} \\
|W| (U - V_3 - E_i)  \\
|W|^2 e^{i \varphi_0} \ear \RB, \label{eqn:eigenstates}
\end{equation}
where $W = \f{1}{\sqrt{2}}(V_1 - i V_2) = \Lb W \Rb e^{-i \varphi_0}$ and $N_i$ is the normalization factor.  For completeness, we have recorded here the eigenenergies and eigenstates for general $\vec{V}$.  The instantaneous eigenenergies for $U=1$, $V_1=-2$ and $V_3 = t$ are shown in Fig.~\ref{fig:rel:AGJ:a1U1}. 

\subsection{\label{ssec:model:IONR} Instantaneous occupation number relaxation approximation}

The wave function of any two-electron spin-singlet state can be factored into a symmetric spatial function and a singlet spin function.
Thus, it is sufficient to consider the \textit{spatial} one-matrix (hereafter, just one-matrix) defined as
\begin{equation}
\gamma_{ij} = \sum_{\sigma} \gamma(i\sigma,j\sigma). 
\label{eqn:one-matrix:spatial}
\end{equation}
In the present model, the one-matrix is a Hermitian $2\times 2$ matrix that can be represented by a Bloch pseudospin vector $\vec{\gamma} = (\gamma_1,\gamma_2,\gamma_3)$ according to $\gamma = I + \vec{\gamma}\cdot \vec{\s}$.  The pseudospin vector $\vec{\gamma}$ should not be confused with the pseudospin vector of a two-level system, because $|\vec{\gamma}|$ is time dependent and generally different than $1$, while the modulus of the latter is always equal to $1$, i.e., it remains on the Bloch sphere.  The natural orbitals in the site basis are expressed as 
\begin{align}
\phi_a &= \LB \bar{c} \cos(\theta/2) e^{-i\varphi/2} \\ \sin(\theta/2) e^{i\varphi/2} \ear \RB \nonumber \\ 
\phi_b &= \LB \bar{c} -\sin(\theta/2) e^{-i\varphi/2} \\ \cos(\theta/2) e^{i\varphi/2} \ear \RB, \label{eqn:nat:orb}
\end{align}
and the occupation numbers are $f_a = 1+A$ and $f_b = 1-A$ with $A = |\vec{\gamma}|$.  In spherical coordinates, $\vec{\gamma} = A (\sin \theta \cos \varphi, \; \sin \theta \sin \varphi, \; \cos \theta)$.  The $\gamma_1$ component is proportional to the kinetic energy, while $\gamma_2$ can be interpreted roughly as an analog of current.  Since $\gamma_3$ is local in the site basis, it represents the density variable.

\subsubsection{\label{sssec:model:IONR:EOM} Equation of motion}

For the present model, the one-matrix EOM becomes
\begin{align}
i \partial_t \gamma = [h, \gamma] + u, \label{eqn:BBGKY:HM}
\end{align}
where $\gamma$, $h$ and $u$ are $2\times 2$ matrices and $h = \vec{V} \cdot \vec{\sigma}/2$.  The inhomogeneous term $u$, which depends on the two-matrix (see Eq.~\ref{eqn:veff}), embodies the contribution of electron-electron interactions.  In the IONR approximation, the exact two-matrix functional is approximated by the adiabatic extension of the ground-state functional $\Gamma[\gamma]$.  For two-electron systems in spin-singlet states, an exact expression for the ground-state wave function in terms of natural orbitals and occupation numbers is known\cite{loewdin1955} up to sign factors that should be chosen to give the absolute minimum of the energy.\cite{kutzelnigg1963}  In the present model, $\Lb\Psi\Rr = \sqrt{f_a/2} \Lb\Phi_{aa}\Rr + \eta \sqrt{f_b/2} \Lb\Phi_{bb}\Rr$, where $\Lb\Phi_{ii}\Rr = \hat{a}_{i\uparrow}^{\dag} \hat{a}_{i\downarrow}^{\dag} \left|0\right>$ and $\eta$ is a sign factor that is equal to $-1$.  Here, $\hat{a}_{i\s}^{\dag}$ and $\hat{a}_{i\s}$ are the creation and annihilation operators for natural spin-orbital $\phi_{i\s}$ ($i=a,b$).  Therefore, the exact ground-state reconstruction $\Gamma[\gamma] = \Lb\Psi\Rr \Ll\Psi\Rb$ is known and can be used to approximate $u$.  In the natural orbital basis, the off-diagonal elements of $u$ are found to be
\begin{align}
u_{ab} = -u_{ba}^* &=U (1 + \cos \beta) \sin \theta \cos \theta, \label{eqn:wab}
\end{align}
where $\beta = \sin^{-1} A$.  To calculation $u_{ab}$, it is convenient to use the following expression, which follows from Eq.~(\ref{eqn:veff}): $u_{ij} = 2\sum_{\s} \Ll \Psi \Rb \big[ \hat{a}_{j\s}^{\dag} \hat{a}_{i\s}, \hat{U} \big] \Lb \Psi \Rr$.
The factor $\cos\beta$ in Eq.~(\ref{eqn:wab}) represents the occupation number dependence, while $\theta$ and $\varphi$ represent the dependence on the natural orbitals.  In the adiabatic extension approximation, the diagonal elements of $u$ are identically zero even though we are using the \textit{exact} ground-state reconstruction.  As a result, the adiabatic extension approximation predicts time-independent occupation numbers (cf.~Eq.~\ref{eqn:occnum:evolution}).  This deficiency is corrected below in the IONR approximation.  As an aside, it is interesting to examine how time-dependent occupation numbers are generated in the exact equation of motion.  The exact time-dependent wave function can be expressed as $\Lb\Psi\Rr = e^{i \psi/2} \sqrt{f_a/2} \Lb\Phi_{aa}\Rr - e^{-i \psi/2} \sqrt{f_b/2} \Lb\Phi_{bb}\Rr$, where the only difference with respect to the ground-state wave function is the relative phase factor between the terms.  If this expression is used to calculate the exact $u$, one obtains $u_{ab} = -u_{ba}^*= U (1 + e^{i \psi} \cos \beta) \sin \theta \cos \theta$ and $u_{aa} = - u_{bb} = i U \cos \beta \sin^2 \theta \sin \psi$.  The exact $u$ generates time-dependent occupation numbers because its diagonal components are nonzero when there is a nontrivial relative phase between the different configurations (Slater determinants) that comprise the wave function.  However, no functional approximations are known for the relative phases. 

The following EOM for the pseudospin vector $\vec{\gamma}$, which follows from Eq.~(\ref{eqn:BBGKY:HM}), provides a geometric interpretation for the conservation of the occupation numbers in the adiabatic extension approximation:   
\begin{align}
\partial_t \vec{\gamma} &= \vec{V}\times \vec{\gamma} + \vec{U},  \label{eqn:BBGKY:Bloch}
\end{align}
where $\vec{U}$ is defined by $u = \vec{U} \cdot \vec{\sigma}$.  Equation (\ref{eqn:BBGKY:Bloch}) is similar to the Landau-Lifshitz-Gilbert equation or the Bloch 
\begin{figure}[t!]
\includegraphics[width=0.8\columnwidth]{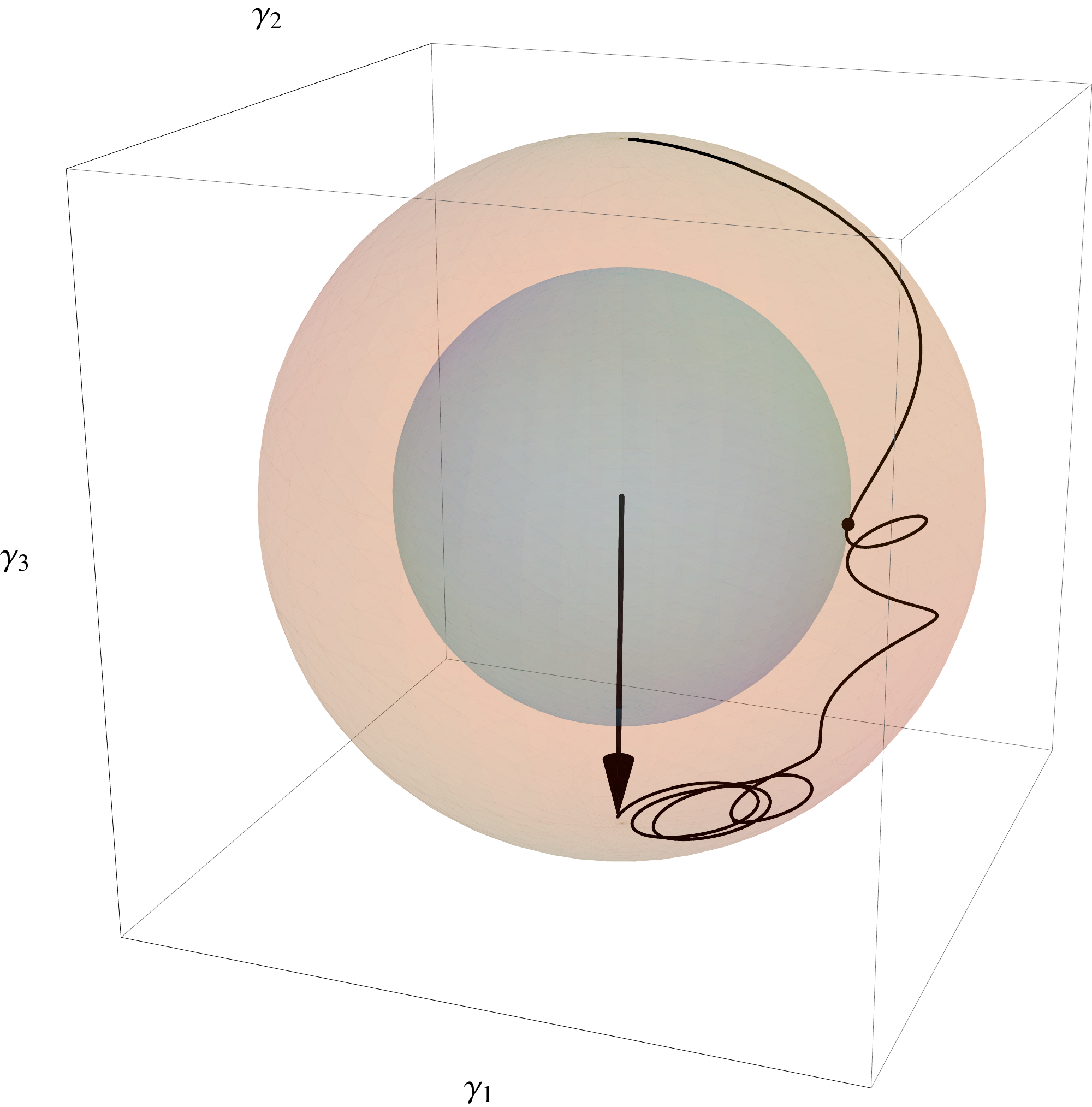}
\caption{\label{fig:trajectory} Bloch sphere and trajectory of the pseudospin $\vec{\gamma}$.  Linear-time potential $\vec{V} = (-2,0,3t)$, $U = \f{7}{2}$, $t \in(-\infty,4.7)$.  The point of intersection shows $\vec{\gamma}$ at the time $t\approx -0.261$, where $|\vec{\gamma}|$ is minimum; the inner sphere shows $|\vec{\gamma}|_{\rm min}\approx 0.633$.}
\end{figure}  
equation with dissipation.  The inhomogeneous term $\vec{U}$ is responsible for changing the modulus of $\vec{\gamma}$ (Fig.~\ref{fig:trajectory}), which corresponds to changing the occupation numbers.  However, in the adiabatic extension approximation, $\vec{U}$ is always perpendicular to $\vec{\gamma}$ so that Eq.~(\ref{eqn:BBGKY:Bloch}) preserves $\Lb \vec{\gamma} \Rb$.  

\subsubsection{\label{sssec:model:IONR:Schroedinger} Effective Schr\"odinger equation}

In the IONR approximation, $A(t)$ is determined from the instantaneous minimization of the ground-state energy functional $E_v[\gamma]$ subject to the constraint that $\theta$ and $\varphi$ are equal to $\theta(t)$ and $\varphi(t)$, the values obtained from the effective Schr\"odinger equation, Eq.~(\ref{eqn:orb:evolution}).  For the present model, 
\begin{align}
E_v[\gamma] = \vec{V}\cdot\vec{\gamma} + U \sin^2 \f{\beta}{2} + U \cos^2 \f{\beta}{2} \cos^2 \theta,
\end{align}
and the value of $\beta = \sin^{-1} A$ that minimizes the energy for given $\theta$ and $\varphi$ is 
\begin{equation}
\tilde{\beta} = - \cot^{-1} \LB \f{U}{2} \f{\sin^2 \theta}{\vec{V}\cdot \hat{\gamma}} \RB. \label{eqn:beta:min}
\end{equation}
The effective Schr\"odinger equation for the orbitals, Eq.~(\ref{eqn:orb:evolution}), becomes
\begin{equation}
i \partial_t \phi_i = \LB \hat{t} + \hat{v}_{\mathrm{eff}} \RB \phi_i, \label{eqn:Schroed:eff}
\end{equation}
where $\hat{t} = V_1 \sigma_1/2$ and $\hat{v}_{\mathrm{eff}} = \hat{v} + \hat{v}_{ee}$ with $\hat{v} = V_3 \sigma_3/2$ and $\hat{v}_{ee} = \vec{V}_{ee}\cdot \vec{\sigma}/2$.  In spherical coordinates, $\vec{V}_{ee}$ has components
\begin{align}
V_{ee,\theta} &= - U \cot \frac{\tilde{\beta}}{2} \sin \theta \cos \theta \nonumber \\
V_{ee,\varphi} &= 0.
\end{align}
The radial component of $\vec{V}_{ee}$ (the component parallel to $\vec{\gamma}$), which corresponds to the ``diagonal'' component $v_{ee,aa} - v_{ee,bb}$, is nonunique as discussed in Sec.~\ref{ssec:dyneqns}.  We remark in passing that $\hat{v}_{ee}$, apart from this nonunique component, is equal to $\delta W[\gamma]/\delta \gamma$, where $W[\gamma]$ is the ground-state electron-electron interaction energy functional.  Therefore, in the IONR approximation, $\hat{v}_{ee}$ is the adiabatic extension of the electron-electron part of the \textit{generalized} ground-state KS potential.\cite{requist2008}

\subsubsection{\label{sssec:model:IONR:linearization} Linearization}

It is interesting to examine the role of the nonlinearity of $\vec{V}_{ee}[\gamma]$ by carrying out a linearization and semilinearization with respect to a reference dynamics or zeroth-order dynamics.  A suitable reference is the instantaneous ground-state one-matrix $\gamma^{(0)}(t)$, from which the exact one-matrix does not deviate too greatly in the adiabatic regime.  Thus, the lowest-order (linear) approximation consists in solving Eq. (\ref{eqn:Schroed:eff}) with $\vec{V}_{ee}$ evaluated at $\theta = \theta^{(0)}$ and $\varphi = \varphi^{(0)}$, where $\theta^{(0)}$ and $\varphi^{(0)}$ are the angular variables obtained from $\gamma^{(0)}$ ($\varphi^{(0)}=0$ if $V_2=0$).  In the next-lowest order (semilinear) approximation, we solve Eq. (\ref{eqn:Schroed:eff}) self-consistently with the potential  
\begin{align}
\vec{V}^{\mathrm{sl}} &= \vec{V}^{\mathrm{eff}}[\gamma^{(0)}] + \left.\f{\partial \vec{V}^{\mathrm{eff}}}{\partial \cos \theta}\right|_{\gamma^{(0)}} \delta (\cos \theta) \nonumber \\
&+ \left.\f{\partial \vec{V}^{\mathrm{eff}}}{\partial \cos \varphi}\right|_{\gamma^{(0)}} \delta (\cos \varphi)
\end{align}
where, for example, 
\begin{align}
\delta (\cos \theta) &= \cos \theta - \cos \theta^{(0)} \nonumber \\
&= (|a_1|^2 - |a_2|^2) - (|a^{(0)}_1|^2 - |a^{(0)}_2|^2),
\end{align}
where $a_i$ are the elements of $\phi_a$.  We have linearized in the variables $\cos \theta$ and $\cos \varphi$ rather than $\theta$ and $\varphi$ because $\cos \theta$ is closely related to the density variable $\gamma_3 = A \cos \theta$, which, in the following section, will be used in the linearization of the TD KS equations.  The linear and semilinear approximations are compared with the full IONR approximation in Sec.~\ref{sssec:validity}.

\subsection{\label{ssec:model:AEA} Adiabatic extension in time-dependent density functional theory}

To better understand the nature of the IONR approximation, it is helpful to compare it with ADFT.  The methods are similar because in both cases the orbitals are determined by an effective Schr\"odinger equation, in which the effective potential is the adiabatic extension of a ground-state effective potential.

\subsubsection{\label{sssec:model:AEA:equations} Time dependent Kohn-Sham equations}

For the two-site Hubbard model, the TD KS equations in the adiabatic extension approximation are
\begin{align}
i \partial_t \phi_a &= \LB \hat{t} + \hat{v}^{ae}\RB \phi_a, \nonumber \\
\gamma_3 &= 2 \big< \phi_a \big| \f{1}{2} \hat{\sigma}_3 \big| \phi_a \big>, \label{eqn:TDKS:HM}
\end{align}
where the factor of $2$ comes from the fact that the orbital $\phi_a$ is doubly-occupied and $\hat{v}^{ae} = (V_{ae}/2) \hat{\sigma}_3$ with
\begin{align}
V_{ae}(t) &= V(t) + V_H(t) + \left.\f{\partial E_{xc}}{\partial \gamma_3}\right|_{\gamma_3 = \gamma_3(t)} \nonumber \\
&= -V_1 \f{\gamma_3(t)}{\sqrt{1 - \gamma_3(t)^2}} \nonumber \\
&= -V_1 \cot \left[\theta_s(t)\right], 
\end{align}
where $V_H(t)$ is the time-dependent Hartree potential and $E_{xc}$ is the ground-state xc energy functional.  The adiabatic KS energies are $\epsilon_{a,b} = \mp (V_1/2) \csc \theta_s$.  The density variable in this model is $\gamma_3 = \cos \theta_s$, where $\theta_s$ should not be confused with $\theta$ defined in Sec.~\ref{ssec:model:IONR}.

\subsubsection{\label{sssec:model:AEA:linearization} Linearization}

In analogy with Sec.~\ref{sssec:model:IONR:linearization}, we carry out a linearization and semilinearization with respect to the instantaneous ground-state density $\gamma_3^{(0)}(t)$. The lowest-order (linear) approximation consists in evolving Eq.~(\ref{eqn:TDKS:HM}) with $V^{(0)}_{ae} = V_{ae}(\gamma_3^{(0)})$.  In the next-lowest order (semilinear) approximation, we evolve Eq.~(\ref{eqn:TDKS:HM}) with
\begin{align}
V_{ae}^{sl}(\gamma_3) &= V_{ae}(\gamma_3^{(0)}) + \left.\f{\partial V_{ae}}{\partial \gamma_3}\right|_{\gamma_3^{(0)}} \delta \gamma_3 \nonumber \\
&= V_{ae}(\gamma_3^{(0)}) + \LB \chi_s^{-1} - \chi^{-1} \RB \delta \gamma_3 
\end{align}
where $\chi_s$ and $\chi$ are the instantaneous static ground-state response functions.  Thus, the better $\chi_s$ approximates $\chi$, the less significant is the nonlinearity.  The linear deviation of the density is $\delta \gamma_3 = |(a_1|^2 - |a_2|^2) - \big( |a^{(0)}_1|^2 - |a^{(0)}_2|^2 \big)$, where $a_i$ are the elements of the KS orbital $\phi_a$.  The TD KS equations become
\begin{align}
i \partial_t \LB \bar{c} a_1 \\ a_2 \ear \RB = \big[ \hat{t} + \f{1}{2} (V^{(0)}_{ae} - C \delta \gamma_3) \sigma_3\big] \LB \bar{c} a_1 \\ a_2 \ear \RB, 
\label{eqn:nonlinear:LZ}
\end{align}
where $C = \chi_s^{-1} - \chi^{-1}$.  Although the potential has been linearized with respect to $\delta \gamma_3$, the equations remain nonlinear in the amplitudes $a_i$.  This nonlinearity is essential
\begin{figure*}[t!]
\centering
\includegraphics[width=2\columnwidth]{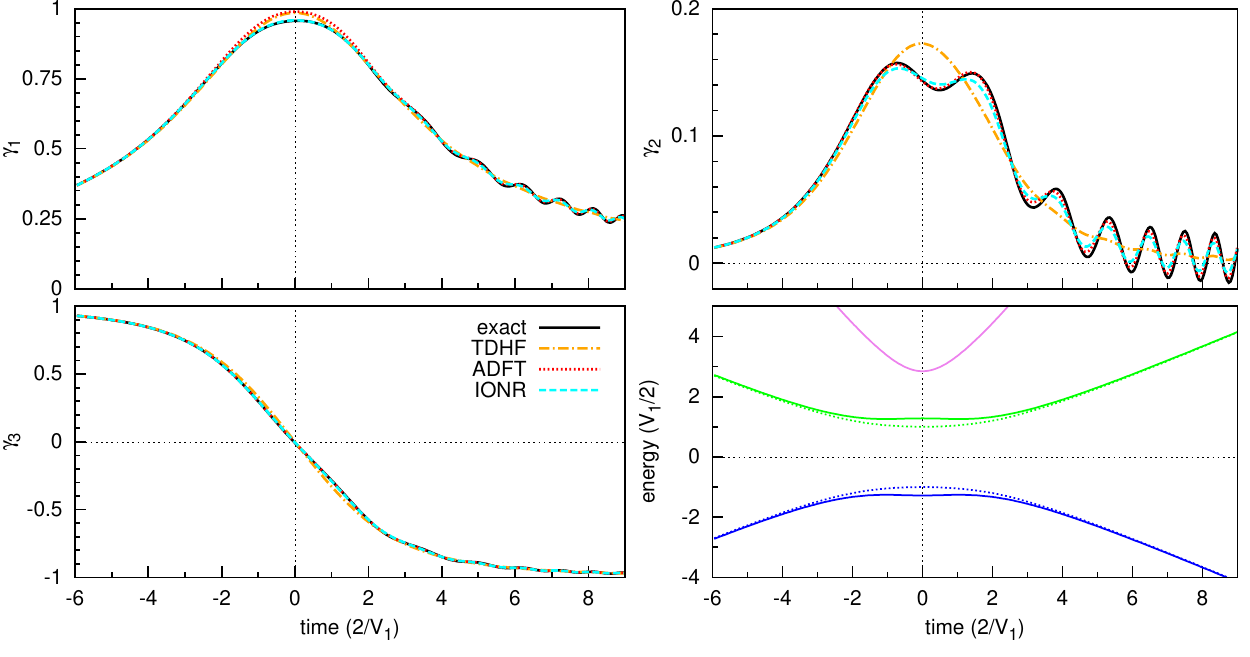} 
\caption{\label{fig:comp:AGJ:a1U1} (Color online) Time-dependence of the one-matrix.  Linear potential, $\tau = 1$, $U = 1$.  Top left, top right, and bottom left: $\gamma_1$, $\gamma_2$, and $\gamma_3$.  Bottom right: adiabatic many-body energies [solid lines] and adiabatic KS energies [dotted lines].}
\end{figure*} 
for the accurate description of nonadiabatic effects.  Equation (\ref{eqn:nonlinear:LZ}) has the form of a nonlinear LZ problem.  In fact, similar equations have been studied as simple models of the Gross-Pitaevskii equation in the context of trapped Bose-Einstein condensates.\cite{milburn1997,smerzi1997,wu2000}  The linear and semilinear approximations are compared with the full ADFT in Sec.~\ref{sssec:validity}.

\subsection{\label{ssec:numerical simulations} Numerical simulations}

We have performed simulations comparing the IONR approximation, TDHF, ADFT and the numerically exact solution.  The following two types of time-dependent external potentials were considered:  a linear potential $V_3 = t/\tau$ and a pulse potential $V_3 = \delta/\cosh(t/\tau)$.  The case of the linear potential can be interpreted as the scattering of two species --- one binding two electrons and one that can accept electrons.  We are interested in the dynamics of systems that start in the ground state, and in principle we can take any time as the initial time $t_0$.  Since for both potentials $t = -\infty$ is a point where the nonadiabatic coupling vanishes and the instantaneous eigenenergies are nondegenerate, it is intuitively clear that $\gamma(t)$ will approach a unique time-dependent function as we let $t_0$ tend to $-\infty$, always choosing the ground state for the initial state.  In our simulations, we set $t_0 = -p$ and choose larger and larger values of $p$ until $\gamma(t)$ is sufficiently converged for all $t \geq t_0$.  We discuss results obtained for the linear potential in Secs.~\ref{sssec:nonadiabatic}-\ref{sssec:validity} and for the pulse potential in Sec.~\ref{sssec:pulse}.

\subsubsection{\label{sssec:nonadiabatic} Nonadiabatic effects}

\begin{figure*}[t!]
\includegraphics[width=2\columnwidth]{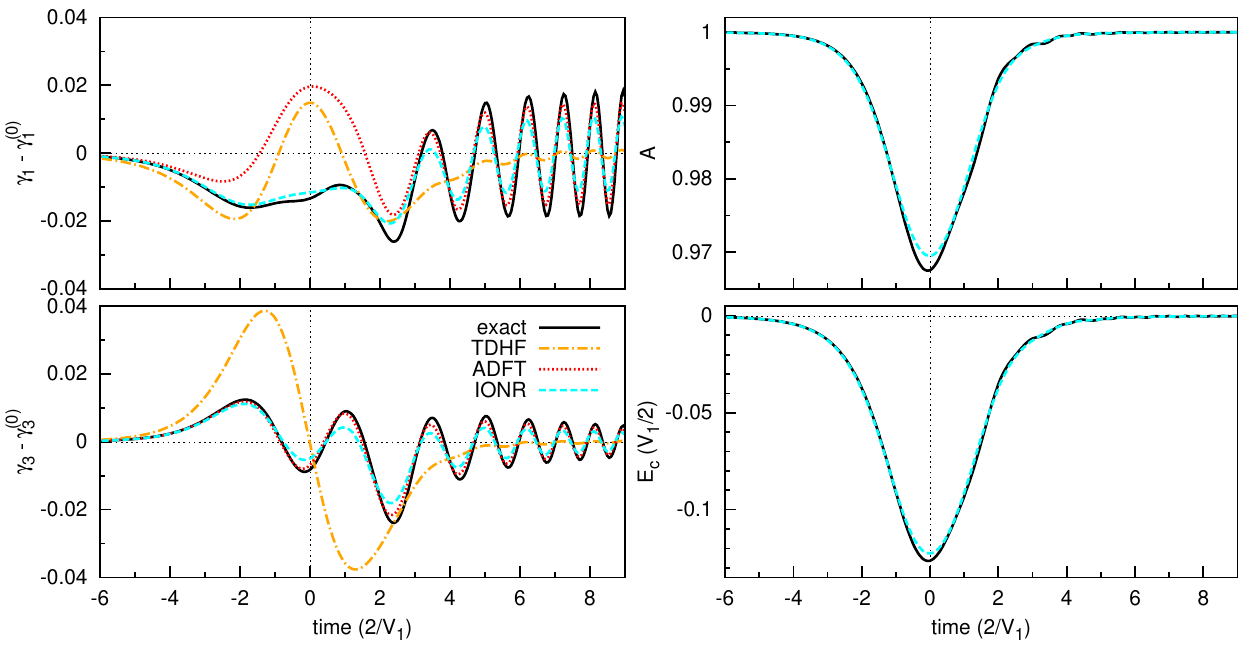}
\caption{\label{fig:rel:AGJ:a1U1} (Color online) Time dependence of the one-matrix relative to the instantaneous ground-state one-matrix.  Linear potential, $\tau = 1$, $U=1$.  Top left and bottom left: $\gamma_1 - \gamma^{(0)}_1$ and $\gamma_3 - \gamma^{(0)}_3$.  Top right: $A$.  Bottom right: correlation energy $E_c$.}
\end{figure*}

The time-dependence of the one-matrix is plotted in Fig.~\ref{fig:comp:AGJ:a1U1} for $U=1$, $V_1/2=-1$ and $V_3 = t$ ($\tau=1$).  For these parameters, the system is in an adiabatic regime and, as seen in the plot of $\gamma_3$, both electrons are transferred smoothly from site $1$ to site $2$ ($\gamma_3 = 1\rightarrow -1$). The model has two dimensionless parameters: $U/V_1$ and $V_1 \tau$.  In all of the following, we take the hopping parameter $V_1/2$ to be equal to $-1$, and refer to $U$ and $\tau$ as dimensionless parameters.  Figure~\ref{fig:rel:AGJ:a1U1} shows the deviation of the one-matrix in Fig.~\ref{fig:comp:AGJ:a1U1} from the instantaneous ground-state one-matrix $\gamma^{(0)}(t)$.  We can identify two principal nonadiabatic effects:  1) coherent oscillations in $\gamma(t)$ as $t \rightarrow \infty$ and 2) mixing with excited states near avoided crossings of the adiabatic energy levels.  The asymptotic oscillations originate from nonadiabatic (LZ) transitions.  These are transitions between the initial and final ($t\rightarrow \pm \infty$) adiabatic eigenstates and generally leave the system in a nonstationary state.  The admixture of excited states near avoided crossings is apparent in the deviation of $\gamma_2$ from $\gamma^{(0)}_2$ ($\gamma^{(0)}_2=0$ for all time) and $A$ from $A^{(0)}$.  The projection of the wave function onto an excited state can be quite large near an avoided crossing, even if ultimately, for large times, the amplitude becomes very small.  The right panels in Fig.~\ref{fig:rel:AGJ:a1U1} show that the instantaneous correlation energy, $E_c = E - E_{HF}$, is strongly correlated with the deviation of the occupation numbers from $0$ or $1$.

\begin{figure*}[ht!]
\includegraphics[width=2\columnwidth]{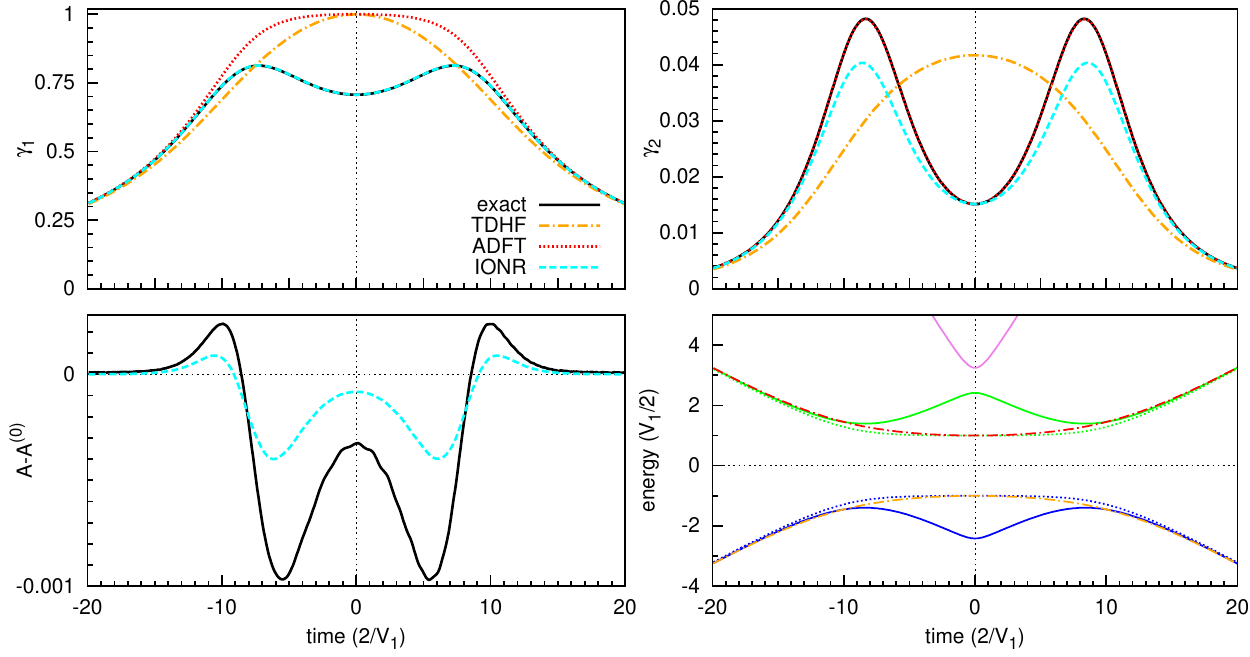} 
\caption{\label{fig:double:AGJ:a05U4} (Color online)  Linear potential, $\tau = 2$, $U = 4$.  Top left and top right:  $\gamma_1$ and $\gamma_2$.  Bottom left: $A - A^{(0)}$.  Bottom right: adiabatic energy levels (colors same as Fig.~\ref{fig:comp:AGJ:a1U1} with the addition of HF eigenvalues [dashed lines]).}
\end{figure*}

The interplay between electron-electron interactions and nonadiabatic effects is, in leading order, mediated by the adiabatic energy levels, which will be discussed in detail below.  Figure~\ref{fig:double:AGJ:a05U4} shows the one-matrix in a strongly interacting case; $\tau = 2$ and $U=4$.  Stronger interactions lead to more pronounced curvature in the energy level profiles (compare Figs.~\ref{fig:comp:AGJ:a1U1} and \ref{fig:double:AGJ:a05U4}).  Interestingly, the adiabatic KS and HF energies do not display such curvature.
 
In Fig.~\ref{fig:trajectory}, we have plotted the trajectory of the pseudospin vector $\vec{\gamma}$ for $U = 7/2$ and $\tau = 1/3$.  The vector starts at the north pole at $t=-\infty$ and, in following the driving vector $-\vec{V}$, it rotates toward the south pole.  Due to the finite probability of nonadiabatic transition, it does not reach the south pole and instead spirals around it perpetually.  Near the equatorial plane of the sphere ($t \sim 0$), the occupation numbers deviate significantly from their original values [$f_a(-\infty) = 2$ and $f_b(-\infty) = 0$] so that $\vec{\gamma}$ leaves the surface of the Bloch sphere ($|\vec{\gamma}|$=1).  Remarkably, the trajectory exhibits a fairly regular spiraling pattern even at such times.  In contrast to the persistent spiraling as $t \rightarrow \infty$, for times near the avoided crossings the spiraling is not tangential to the surface of the Bloch sphere.

\subsubsection{\label{sssec:asymptotic:oscillations} Asymptotic oscillations}

By studying the amplitude and phase of the asymptotic oscillations, we can extract information about the nonadiabatic transitions.  For the range of parameters investigated here, the asymptotic oscillations are dominated by a single time-dependent frequency $\Omega(t)=E_2(t)-E_1(t)$.  Thus, the one-matrix is well described by the expression
\begin{align}
\gamma_i = \overline{\gamma}_i + \Delta_i \cos\left(\int_0^t dt\pr \Omega(t\pr) - \Theta_i \right). \label{eqn:asym:oscillations}
\end{align}
The quantities $\overline{\gamma}_i$, $\Delta_i$ and $\Theta_i$ are related to the final amplitudes of the adiabatic states.  Let us write the many-body wave function as
\begin{align}
\Lb\Psi(t)\Rr = \sum_{k=1}^3 c_k(t) e^{-i \int_0^t dt\pr E_k(t\pr)} \Lb \psi_k(t)\Rr, \label{eqn:expansion}
\end{align}
where $\Lb \psi_k \Rr$ and $E_k$ are the instantaneous eigenstates and eigenenergies given in Eqs. (\ref{eqn:energies}) and (\ref{eqn:eigenstates}).  Except for low values of $\tau$ ($\tau \lesssim 1/4$), the final amplitude of the highest energy adiabatic state $\Lb\psi_3\Rr$ is much less than those of the lowest two states because it is separated from the ground adiabatic state by a larger energy gap.  Therefore, as a first approximation in the adiabatic regime, let us truncate the expansion in Eq.~(\ref{eqn:expansion}) to the lowest two levels.  Then, we find
\begin{align}
\overline{\gamma}_i &= |c_1|^2 \Ll \psi_1 \Rb \f{1}{2} \hat{\sigma}_i \Lb \psi_1 \Rr + |c_2|^2 \Ll \psi_2 \Rb \f{1}{2} \hat{\sigma}_i \Lb \psi_2 \Rr \nonumber \\
\Delta_i &= 2 |c_1| |c_2| \Ll \psi_1 \Rb \f{1}{2} \hat{\sigma}_i \Lb \psi_2 \Rr \nonumber \\
\Theta_i &= \mathrm{Arg}(c_2/c_1). \label{eqn:oscillations} 
\end{align}
where here $c_1$, $c_2$, $\Lb \psi_1 \Rr$ and $\Lb \psi_2 \Rr$ are evaluated at $t=\infty$.  In Fig.~\ref{fig:scattering}, the amplitude $|\Delta_1|$ and phase $\Theta_1$ obtained from
fitting the simulations to Eq.~(\ref{eqn:asym:oscillations}) are shown as functions of $\tau^{-1}$, which can be interpreted as a scattering ``velocity.''  The amplitude of the oscillations is found to be quite sensitive to $\tau^{-1}$, even ``collapsing'' for particular values.  Similarly, if $|\Delta_1|$ is plotted with respect to $U$ for fixed $\tau^{-1}$, it is seen that there is a series of values of $U$ for which the oscillations collapse.  In Sec. \ref{sssec:ICA}, it will be shown that both of these instances of collapse are manifestations of the same phenomenon, namely an interference effect often referred to as Stueckelberg oscillations\cite{stueckelberg1932} or Landau-Zener interferometry.  Here, \textit{oscillations} refers to the fact that the effect is quasiperiodic and mediated by a sine factor.  Stueckelberg oscillations are observable in a variety of settings (see for example Refs. \onlinecite{berns2006} and \onlinecite{berns2008}).  As far as we can ascertain within the numerical precision of the simulations, the amplitude of the oscillations collapses all the way to zero in all of the approximations investigated --- TDHF, ADFT and IONR.  In contrast, in the numerically exact solution the amplitude of the oscillations reaches instead a finite minimum value. 

The results in Fig.~\ref{fig:scattering}, taken as a whole, indicate that the IONR approximation and ADFT perform comparably.  The IONR approximation gives more accurately the critical values of $\tau^{-1}$ for which the oscillations collapse.  Although in principle TDDFT need only give exactly the diagonal element $\gamma_3$ --- the density variable in the present model --- and not the off-diagonal elements $\gamma_1$ and $\gamma_2$, it is nevertheless appropriate to compare the methods on the basis of the asymptotic oscillations in $\gamma_1$ because these oscillations have the same origin as corresponding oscillations in $\gamma_3$, which, however, have an amplitude that decays to zero as $t \rightarrow \infty$ due to the fact that $\lim_{t\rightarrow \infty} \langle \psi_1 | \hat{\sigma}_3 | \psi_2 \rangle = 0$.  The phase $\Theta_1$ exhibits a striking resonance behavior with respect to $\tau^{-1}$, jumping by $\pi$ as it passes through $\pm \pi/2$.  The resonances coincide with the minima of $|\Delta_1|$.  In the elucidation of this resonance phenomenon, it will be helpful to have a method for estimating the asymptotic final values of $c_1$ and $c_2$ that appear in Eq.~(\ref{eqn:oscillations}).  We shall now describe such a method.

\subsubsection{\label{sssec:ICA} Independent crossing approximation}

The final amplitudes $c_i(\infty)$ of the adiabatic states are related to the initial amplitudes $c_i(-\infty)$ by a unitary scattering matrix.  In the adiabatic regime, the scattering matrix can often be calculated with good accuracy in the so-called independent crossing approximation (ICA).\cite{brundobler1993}  Consider a multilevel system in which the adiabatic energy levels undergo pairwise avoided crossings.  In the adiabatic limit, nonadiabatic transitions are typically localized near the avoided crossings and each avoided crossing can be described by a scattering matrix that connects the amplitudes of its incoming and outgoing adiabatic states.  Between adjacent avoided crossings, the evolution is nearly adiabatic and the components of the wave function simply acquire dynamical and geometric phases (there is no geometric phase in our model because $\vec{V}$ is in the $xz$-plane and does not encircle the origin; however, it will appear when $\vec{V}$ in fully 3-dimensional and subtends a nonzero solid angle).  An approximation to the scattering matrix for the full time interval can be constructed by taking the time-ordered product of the individual scattering matrices, interposed by diagonal matrices describing the dynamical and geometric phases, provided such a time-ordering is possible.  In the case of the linear-time external potential, the ground adiabatic state undergoes two avoided crossings with the first excited adiabatic state (see Fig.~\ref{fig:double:AGJ:a05U4}).  These crossings are related by reflection about $t=0$.  Although the adiabatic energy levels do not intersect for real time, there are generally crossing points in the complex time plane.  For example, for $U=1$ and $\tau=1$, levels 1 and 2 cross at $t \approx \pm 1.2737 \pm i 2.4584$.  Let $t_a$ be the crossing point with $\mathrm{Re} \; t < 0$ and $\mathrm{Im}\; t >0$, and let $t_b$ be the crossing point with $\mathrm{Re} \; t > 0$ and $\mathrm{Im}\; t >0$.  Between the levels $2$ and $3$, there are two complex conjugate crossing points; for $U=1$ and $\tau=1$, they are $t = \pm i 1.0757$.  Let $t_c$ be the crossing point in the upper half plane.  Levels $1$ and $3$ do not cross for any complex time, so direct transitions between these levels are expected to be very weak. 

The scattering matrix for the avoided crossing near $t_a$ can be approximated by
\begin{equation}
S^a = \LB \begin{array}{ccc} R_a & -T_a^* & 0 \\ T_a & R_a^* & 0 \\ 0 & 0 & 1 \end{array} \RB.
\end{equation}
Due to the time reversal symmetry of the adiabatic energy levels and the nonadiabatic coupling $A_{21} = -i\left<\psi_2| \partial_t \psi_1 \right>$ between levels 1 and 2, the corresponding matrix near $t_b$ is $S^b = (S^a)^{\dag}$.  The scattering matrix near $t_c$ is approximated by
\begin{equation}
S^c = \LB \begin{array}{ccc} 1 & 0 & 0 \\ 0 & R_c & -T_c^* \\ 0 & T_c & R_c^* \end{array} \RB.
\end{equation}
In the ICA, the full scattering matrix connecting the adiabatic states at $t=-\infty$ and $t=\infty$ is 
\begin{align}
S &= D^{0b} S^b D^{bc} S^c D^{ca} S^a D^{a0},
\end{align}
where $D^{\alpha \beta}$ are diagonal matrices containing dynamical and geometric phases, for example, $D^{a0}$ has the elements $D^{a0}_{kk} = e^{-i \Re\int_0^{t_a} dt (E_k - i\langle \psi_k | \partial_t \psi_k \rangle)}$.  The full scattering matrix gives an estimate for the asymptotic final amplitudes $c_i^{ICA}=S_{i1}$:
\begin{align}
c_1^{ICA}(\infty) &= |R_a|^2 + R_c |T_a|^2 e^{-i \Phi} \nonumber \\
c_2^{ICA}(\infty) &= - R_a T_a \big( e^{i \Phi/2} - R_c e^{-i \Phi/2} \big) \nonumber \\
c_3^{ICA}(\infty) &= T_a T_c e^{-i \Re\int_0^{t_a} dt\pr E_1} e^{-i\Re\int_{t_a}^{t_c} dt\pr E_2} e^{-i \Re\int_{t_c}^0 dt\pr E_3} \label{eqn:asymptotic:amps} 
\end{align}
where $\Phi = \mathrm{Re}\int_{t_a}^{t_b} dt \:\Omega(t)$.  In the limit $\tau^{-1} \rightarrow 0$, $T_c \rightarrow 0$ and $R_c \rightarrow 1$.  Thus, for low $\tau^{-1}$, $c_3^{ICA}(\infty) \approx 0$, and $c_1^{ICA}(\infty)$ and $c_2^{ICA}(\infty)$ can be approximated by a two-level (TL) approximation 
\begin{align}
c_1^{TL}(\infty) &= |R_a|^2 + |T_a|^2 e^{-i \Phi} \nonumber \\
c_2^{TL}(\infty) &= -i 2 R_a T_a \sin \f{\Phi}{2}. \label{eqn:two:level}
\end{align}
Stueckelberg oscillations are encoded in the sine factor, which describes the collapse of the oscillations and the accompanying phase jumps.  The amplitude of the oscillations $|\Delta_1|$ is proportional to $|c_2(\infty)|$ and, in the two-level ICA, $c_2(\infty)$ vanishes for $\mathrm{Re}\int_{t_a}^{t_b} dt \Omega(t) = 2\pi n$; $n = 1,2,3,\ldots$  In this Bohr-Sommerfeld-like condition, $\Omega(t) = E_2(t) - E_1(t)$ depends implicitly on $U$ and $\tau^{-1}$ (and, generally, on any other parameters of the system that affect the adiabatic energy levels).  Therefore, the Stueckelberg oscillations with respect to $U$ have the same origin as those with respect to $\tau^{-1}$.  The vanishing of $c_2^{TL}(\infty)$ can be interpreted as the destructive interference between the two different ``pathways'' through the energy level diagram that start in level 1 at $t=-\infty$ and end in level 2 at $t=\infty$.  In the first, a nonadiabatic transition occurs near $t_a$, in the second, near $t_b$.  The phase associated with each pathway is the product of three types of phases --- dynamical, geometric and scattering.  In the present model, as already mentioned, the geometric phase is zero.  And, due to the condition $S^b = (S^a)^{\dag}$, the contributions of the scattering phases contained in $S^a$ and $S^b$ cancel out.  
\begin{figure}[t!]
\centering
\includegraphics[width=1.0\columnwidth]{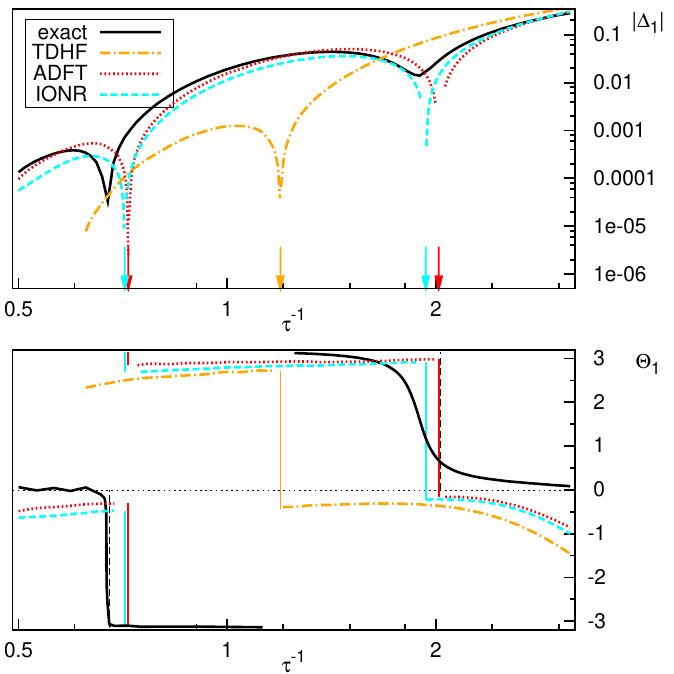} \caption{\label{fig:scattering} (Color online)  $U=1$.  Amplitude $|\Delta_1|$ and phase $\Theta_1$ of the asymptotic oscillations of $\gamma_1$.  Arrows and vertical lines represent the resonances.  Dashed vertical lines represents the resonances predicted by the two-level approximation, Eq. (\ref{eqn:two:level}).}  
\end{figure}

The phase of the oscillations $\Theta_1$ depends on the relative phase of $c_1$ and $c_2$ (here and in the following we omit the argument $t=\infty$).  To the extent that $c_2$ can be approximated by $c_2^{TL}$, we expect to see perfect step-like jumps of $\pi$ in $\Theta_1$, coinciding with the sign changes of $\sin (\Phi/2)$ as it passes through zero.  In Fig.~\ref{fig:scattering}, the jumps predicted in the two-level approximation are indicated by dashed vertical lines.  In contrast to the numerically exact result [solid black line], all of the approximations display discontinuous phase jumps.  Furthermore, the phase in all of the approximations is advanced with respect to the exact result.

\subsubsection{\label{sssec:third} Effect of the third adiabatic state}

The numerically exact solution in Fig.~\ref{fig:scattering} displays a resonance behavior in which the phase increases rapidly but smoothly by $\pi$ at particular values of $\tau^{-1}$.  To understand the broadening of the resonances (note that it is not the nominal amplitude $|\Delta_1|$ but rather the quantity $|\Delta_1|^{-1}$ that becomes very large on resonance), it is necessary to take into account the third adiabatic state $\Lb \psi_3 \Rr$.  The broadening of the resonances is primarily due to transitions to the third state, which destroy the perfect destructive interference that occurs in the two-level approximation.  

We shall now calculate the width of an arbitrary resonance and show that it is proportional to the minimum value of $|\Delta_1|$.  We begin by expressing $c_1^{ICA}$ and $c_2^{ICA}$ as follows:
\begin{align}
c_1^{ICA} &= 1 - |T_a|^2 e^{-i(\Phi-\rho_c)/2} \zeta \nonumber \\
c_2^{ICA} &= -R_a T_a e^{i\rho_c/2} \zeta,\label{eqn:asymptotic:amps:alt}
\end{align}
where $\rho_c = \mathrm{Arg} (R_c)$ and 
\begin{align}
\zeta = e^{i(\Phi-\rho_c)/2} - |R_c| e^{-i(\Phi-\rho_c)/2}
\end{align}
is a key quantity for describing the resonances.  The argument of $\zeta$, 
\begin{align}
\mathrm{Arg} (\zeta) = \tan^{-1}\left[ \frac{1+|R_c|}{1-|R_c|} \tan \LB\f{\Phi-\rho_c}{2}\RB \right],
\end{align}
rises smoothly from $-\pi/2$ to $\pi/2$ near $\Phi-\rho_c = 2\pi n$.  The width (in terms of $\tau^{-1}$) of a resonance at $\tau=\tau\pr$ is defined to be $|d\Theta_1/d(\tau^{-1})|^{-1}_{\tau=\tau\pr}$.  In the following, we shall assume that $|T_a| \ll 1$, which is the case for sufficiently low $\tau^{-1}$.  Hence, the argument of $c_1^{ICA}$ is approximately constant, and the $\tau$-dependence of $\Theta_1$ comes primarily from the argument of $c_2^{ICA}$.  We find 
\begin{align}
\f{d\Theta_1}{d(\tau^{-1})} &= \f{d\mathrm{Arg}(\zeta)}{d(\tau^{-1})} + \f{d\mathrm{Arg}(R_a T_a)}{d(\tau^{-1})} + \f{1}{2} \f{d\rho_c}{d(\tau^{-1})}. \label{eqn:dtheta}
\end{align}
The first term is $\mathcal{O}(\tau^{2})$ because $\Phi \sim \tau$ as $\tau^{-1} \rightarrow 0$, so that the second and third terms, which are $\mathcal{O}(1)$, are negligible.  

To express $d\Theta_1/d(\tau^{-1})$ in terms of $|\Delta_1|$, we first consider the condition 
\begin{align}
0 = \f{d|\Delta_1|}{d(\tau^{-1})} &\approx \f{|\Delta_1||R_c|}{(1-|R_c|)^2} \sin(\Phi-\rho_c) \f{d(\Phi-\rho_c)}{d(\tau^{-1})}, \label{eqn:dDelta1} 
\end{align}
neglecting terms of order $d|c_3|^2/d(\tau^{-1})$.  Equation (\ref{eqn:dDelta1}) gives the resonance condition $\Phi - \rho_c = 2\pi n$.  The phase $\rho_c$, due to scattering with the third level, shifts the resonances with respect to the positions predicted in the two-level approximation.  It becomes significant for moderately large $\tau^{-1}$ (see Fig. \ref{fig:scattering}).  Using Eqs.~(\ref{eqn:asymptotic:amps}) and (\ref{eqn:dDelta1}) in Eq.~(\ref{eqn:dtheta}), we obtain
\begin{align}
\left.\f{d\Theta_1}{d(\tau^{-1})}\right|_{\tau=\tau\pr} = \f{1}{2} \f{1}{|\Delta_1|} \f{|T_c|^2}{|R_c|} \lim_{\tau\rightarrow \tau\pr} \f{d|\Delta_1|/d(\tau^{-1})}{\sin(\Phi-\rho_c)}.
\end{align}
This establishes that the resonance width is proportional to $|\Delta_1|$.  Within the ICA, the minimum value of $|\Delta_1|$ on resonance is $\sqrt{2} |T_a R_a| (1-|R_c|)$, which vanishes very rapidly as $\tau^{-1} \rightarrow 0$ because $|T_a| \rightarrow 0$ and $|R_c| \rightarrow 1$.  Thus, there is an infinite sequence of increasingly sharp resonances in the limit $\tau^{-1} \rightarrow 0$.
\begin{figure}[t!]
\includegraphics[width=\columnwidth]{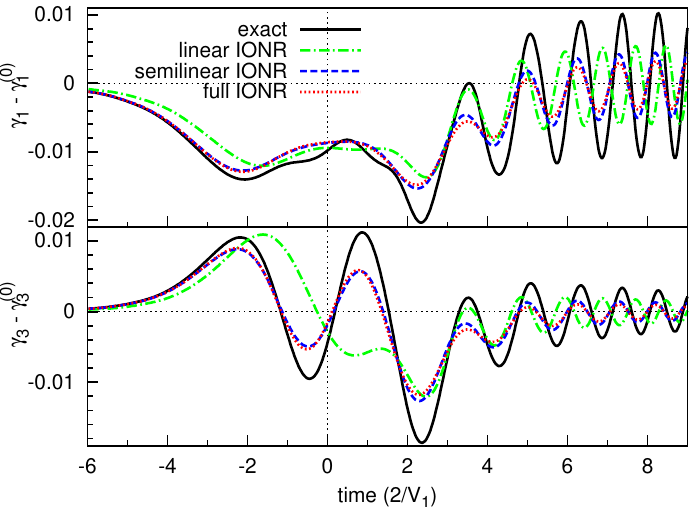} \caption{\label{fig:linearization:AJKL:U125a1} (Color online)  Linearization and semilinearization of the IONR approximation.  Linear potential, $\tau = 1$, $U = 1.25$.}
\end{figure}  

The asymptotic oscillations can be described exactly by including the effect of the third adiabatic state that was neglected in the two-level approximation.  Using $\lim_{t\rightarrow \infty} \left<\psi_i \Lb \hat{\sigma}_1 \Rb \psi_i \right> = 0$ and $\lim_{t\rightarrow \infty} \left<\psi_1 \Lb \hat{\sigma}_1 \Rb \psi_3 \right> = 0$, we find $\overline{\gamma}_1 = 0$ and the following asymptotically exact expression:
\begin{align}
\gamma_1(t) &= |c_1| |c_2| \left<\psi_1 \Lb \hat{\sigma}_1 \Rb \psi_2 \right> \cos \left( \int_0^t dt\pr \phi(t\pr) - U t - \theta_{21} \right) \nonumber \\
&+ |c_2| |c_3| \left<\psi_2 \Lb \hat{\sigma}_1 \Rb \psi_3 \right> \cos \left( \int_0^t dt\pr \phi(t\pr) + U t - \theta_{32} \right),\label{eqn:asym:oscillations:3}
\end{align}
where $c_i$ and $\Lb \psi_i \Rr$ are evaluated at $t=\infty$, $\phi = t/\tau + 2 b^2\tau/t$ and $\theta_{ij} = \textrm{Arg}(c_i/c_j)$.  The third state induces additional oscillations with time-dependent frequency $E_3-E_2 \approx \phi + U$.  If $|c_3|=0$, Eq.~(\ref{eqn:asym:oscillations:3}) reduces to Eq.~(\ref{eqn:asym:oscillations}), while if $|c_3| \approx |c_1|$, the third state generates beats with period $2\pi/U$.  The ICA prediction for the final amplitude of the third state, $c_3 = S_{31} = e^{-2\pi b^2\tau}$, is actually \textit{exact}.  This is a special case of a general result for $n$-level systems with linear time dependence, namely, it has been proved\cite{shytov2004} that the ICA is exact for the scattering matrix elements $S_{1n}$ and $S_{n1}$, where $1$ and $n$ denote the lowest and highest energy adiabatic states.

In the many-body system, the interference of dynamical phases arises from nonadiabatic transitions at two distinct and well separated times ($t_a$ and $t_b$).  Therefore, it is interesting to ask how ADFT and the IONR approximation, which are devoid of memory dependence, are able to describe such an effect.  For ADFT the reason is clear: the same interference phenomenon operates in the KS system.  How nonadiabatic transitions and the interference phenomenon are captured in the generalized KS system represented by Eqs.~(\ref{eqn:orb:evolution}) and (\ref{eqn:occnum:evolution}) is an interesting question for future study.  Both ADFT and the IONR approximation greatly underestimate the resonance width; in fact, within the numerical accuracy of our simulations, we were unable to resolve any of the resonance widths.  This deficiency of the approximations is due to the absence of a spectator KS state (the KS and generalized KS systems have only two states), analogous to the third state in the many-body system, that can ``scatter'' with the states participating in the interference phenomenon.  

It is difficult to calculate the scattering matrix elements in the general case.  In the adiabatic regime the square moduli of the off-diagonal elements, which give transition probabilities, can be estimated with the Dykhne formula,\cite{dykhne1962,davis1976,hwang1977} e.g., for the transition near $t_a$,
\begin{align}
|T_a|^2 = e^{-2 \Im\int_0^{t_a} dt (E_2-E_1)}. \label{eqn:Dykhne}
\end{align}
The Dykhne formula is universal in the sense that it depends only on the adiabatic energies and not the non\-adiabatic coupling.  For the ``reflection'' component of $S^{\alpha}$, we have $|R_{\alpha}| = \sqrt{1 - |T_{\alpha}|^2}$.  However, having only the moduli of $T_{\alpha}$ and $R_{\alpha}$ is not sufficient to fully describe the asymptotic oscillations: the phases are also important.   In Eqs. (\ref{eqn:asymptotic:amps}), we see that the arguments of $R_a$, $T_a$ and $R_c$ directly influence $\Theta_1$ and the location of the resonances.  The $U$ and $\tau$ dependence of these arguments is manifest in the nonvanishing slopes of the plateaus in Fig.~\ref{fig:scattering}.

Progress in the context of time-dependent functional approximations might require confronting memory dependence in an explicit way.  From this point of view, it is noteworthy that the Dykhne formula, as well as the dynamical and geometric phases in $D^{\alpha \beta}$, contain a type of memory dependence due to the time integrals.  The functional dependence enters through the adiabatic energies, which can be taken to be functionals of the instantaneous ground-state one-matrix (ground state density) rather than the true time-dependent one-matrix (time-dependent density).  However, as mentioned above, the Dykhne formula gives only the probability of nonadiabatic transition at an avoided crossing, and the scattering phases are also important.  The nonadiabatic coupling is expected to be important in the calculation of scattering phases.  While the calculation of excited state energies within linear response theory is fairly well developed, much less is known about the nonadiabatic couplings (for example, see Refs. \onlinecite{chernyak2000} and \onlinecite{baer2001}).  

\subsubsection{\label{sssec:validity} Validity of linearization and semilinearization}

The time-dependent Kohn-Sham potential and the effective potential in Eq.~(\ref{eqn:orb:evolution}) are nonlinear functionals of the time-dependent density and time-dependent one-matrix, respectively.  By considering the linear and semilinear versions of these equations, we can investigate the importance of nonlinearity in the description of nonadiabatic effects. 
\begin{figure}[t!]
\includegraphics[width=\columnwidth]{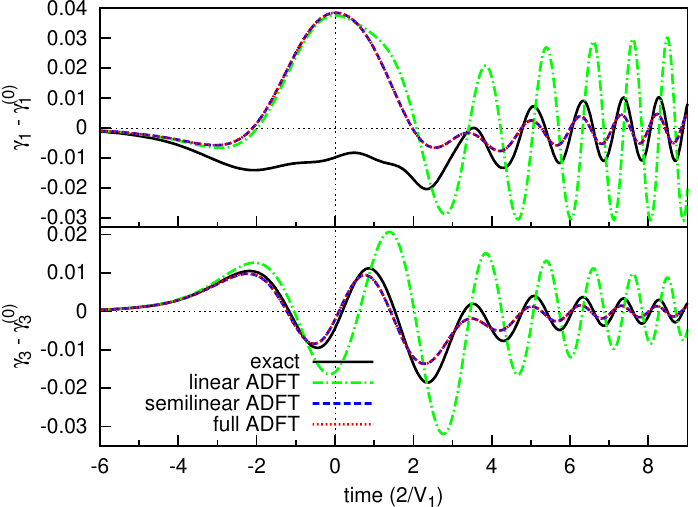}
\caption{\label{fig:linearization:AGHI:U1a1} (Color online) Linearization and semilinearization of ADFT.  Linear potential, $\tau = 1$, $U = 1.25$.}
\end{figure}

The linear and semilinear versions of Eq.~(\ref{eqn:orb:evolution}) in the IONR approximation were introduced in Sec.~\ref{sssec:model:IONR:linearization}.  Figure \ref{fig:linearization:AJKL:U125a1} compares the linear and semilinear versions with the full (unmodified) equations for representative values of $U$ and $\tau$.  Even the linear version generates asymptotic oscillations, although their amplitude, phase and period are generally incorrect.  The period is corrected in the semilinear and full versions.  The semilinear version improves significantly also the amplitude, phase and mean value of the oscillations with respect to the linear version.  

In Sec.~\ref{sssec:model:AEA:linearization}, similar linearizations were performed in ADFT.  Figure \ref{fig:linearization:AGHI:U1a1} compares the linear and semilinear versions with the full equations.  As found in the IONR approximation, the amplitude and phase of the asymptotic oscillations are quantitatively incorrect in the linear version and the nonlinearity of the semilinear version brings them into better agreement with the full version.  
\begin{figure*}[t!]
\includegraphics[width=2\columnwidth]{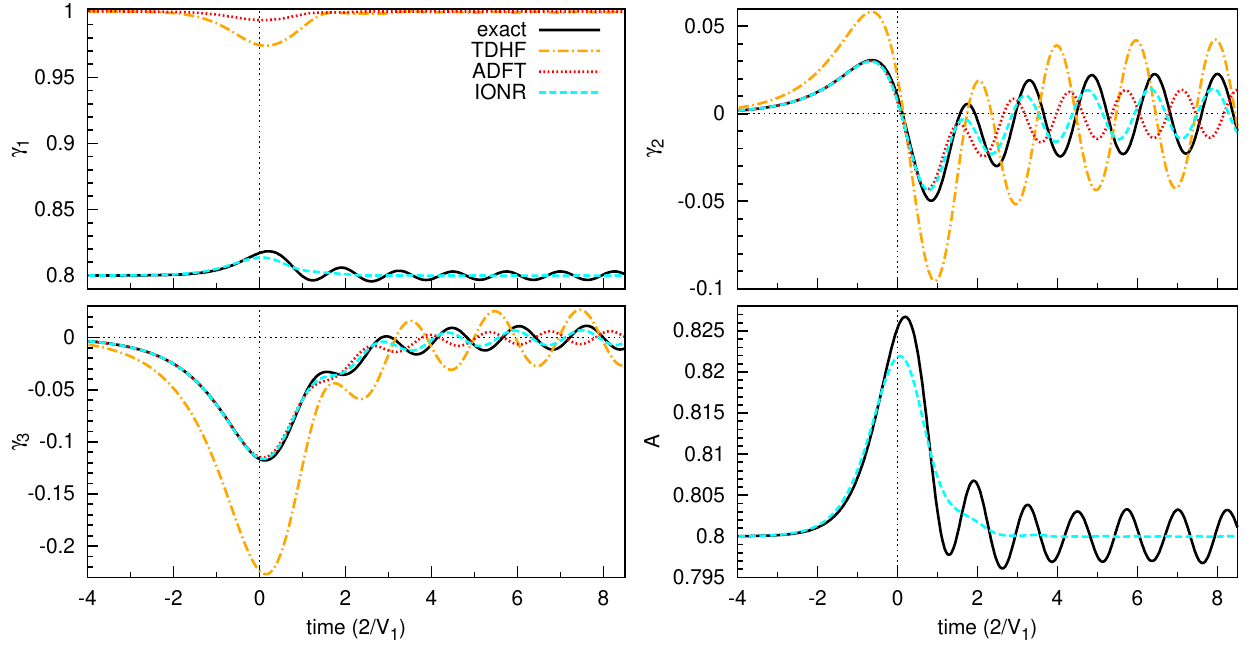}
\caption{\label{fig:pulse:ABG:U3a1d1}  (Color online).  Time dependence of the one-matrix for a pulse potential $V_3 = 1/\cosh t$; $U = 3$.}
\end{figure*}
Surprisingly, only this lowest-order nonlinearity is sufficient to bring the semilinear version into nearly perfect agreement with the full version, except for large values of $U$ and low values of $\tau$ ($U\gtrsim 8$ and $\tau\lesssim \f{1}{8}$).  In contrast to the linear version of Eq.~(\ref{eqn:orb:evolution}), the linear version of the KS equations achieves the correct period because the difference in the adiabatic KS energies $\epsilon_b - \epsilon_a$ equals the difference $E_2-E_1$ in the limit $t \rightarrow \infty$ (see Fig.~\ref{fig:double:AGJ:a05U4}). However, this is a special property of the linear-time potential due to its divergence in the limit $t \rightarrow \infty$.

\subsubsection{\label{sssec:pulse} Pulse potential}

Up to this point, all of the simulations we have reported have used an external potential with linear time dependence.  As such a potential diverges in the limit $|t| \rightarrow \infty$, it has the special property that the ground state at $t = -\infty$, in which both electrons occupy site $1$, is an uncorrelated single Slater determinant. To examine the performance of the approximations in systems with correlated initial states, we have considered additional time-dependent external potentials.  Here we report simulations for a pulse-shaped potential, $V_3 = 1/\cosh(t/\tau)$.

In Figure \ref{fig:pulse:ABG:U3a1d1}, the IONR approximation and ADFT are compared with the numerically exact solution.  The performance of ADFT has worsened for this external potential.  The most significant shortcoming of ADFT is that the period of the asymptotic oscillations is too short.  This is the opposite of what one would expect on the basis of the adiabatic KS energies:  $\epsilon_b-\epsilon_a$ is smaller than $E_2-E_1$, suggesting that the period of the KS oscillations should be too long.  This counterintuitive behavior is accounted for by the nonlinearity of the KS equations.  Although ADFT gives a qualitatively incorrect description of $\gamma_1$, this should not be viewed as a deficiency of the approximation, as TDDFT is only guaranteed to reproduce the density variable $\gamma_3$ and not the full one-matrix.  

The IONR approximation performs better than ADFT for this external potential.  Its most pronounced deficiency is that it does not capture the asymptotic oscillations in $\gamma_1(t)$ and $A(t)$.

\section{\label{sec:conclusions} Conclusions}

Describing strongly-driven electron dynamics from first principles is a challenging problem.  Simulations of a multi-configurational wave function with enough terms to adequately describe correlation have so far been limited to small systems or short simulation times.  Time-dependent density-functional theory is a less computationally demanding approach capable of treating much larger systems.  However, approximations for the xc potential, a complicated nonlocal and memory-dependent functional of the density, must be introduced.  Essentially all calculations to date have employed the adiabatic extension approximation.  Its known deficiencies in linear response calculations may have counterparts in real-time simulations.  Steps toward overcoming the deficiencies of the adiabatic approximation in linear response TDDFT has been made\cite{pernal2007b,pernal2007c,giesbertz2008,giesbertz2009} by using the one-matrix instead of the density as basic variable.  

Working with the one-matrix may have advantages for real-time dynamics as well.  However, applying the adiabatic extension approximation to the one-matrix EOM yields time-independent occupation numbers when any of the available ground-state functionals are used.  We have proposed a simple modification of the adiabatic extension approximation, here called the IONR approximation, in which time-dependent occupation numbers are obtained ``on-the-fly'' from a condition of instantaneous relaxation to the minimum of an adiabatic energy surface.  The IONR approximation has the advantage that it yields time-dependent occupation numbers even for the available functionals, which have proved successful for many ground-state properties.  The motivation for the instantaneous relaxation condition can be understood in light of a sequence of adiabatic energy surfaces that arises from an asymptotic analysis of the many-body Schr\"odinger equation in the limit $\tau \rightarrow \infty$.  Each energy surface obeys an instantaneous minimum principle giving an approximation to the exact one-matrix with error of order $\tau^{-(n+1)}$.  However, arbitrary accuracy cannot be achieved on the basis of this sequence because the existence of nonadiabatic transitions causes the sequence to divergence.

We performed simulations for a model system, which demonstrated that the IONR approximation captures fairly well nonadiabatic effects, such as LZ-type transitions, even though it lacks memory dependence.  Non\-adiabatic transitions leave the system in a nonstationary state, resulting in persistent oscillations in the observables.  The amplitude and phase of the oscillations undergo resonances when either the electron-electron interaction strength $U$ or the characteristic time scale $\tau$ is swept through critical values.  We have found that the IONR approximation describes the persistent oscillations qualitatively correctly over a wide range of $U$ and $\tau$.  It becomes quantitively correct for low values of $U$ and large values of $\tau$.  This is remarkable because the resonance behavior is due to an ultra-nonlocal interference of dynamical phases and scattering phase shifts in the interacting system.  

Although the IONR approximation is able to describe the lowest-order nonadiabatic effects, it has an important limitation: being based solely on the adiabatic extension of ground-state functionals, it does not apply in situations where the time-dependent wave function deviates greatly from the instantaneous ground-state wave function.  Thus, while it may be relevant for adiabatic quantum control problems, it is not applicable to the type of quantum control problem studied in Ref.~\onlinecite{appel2008}, where the target state is an excited state.

\begin{acknowledgments}
This work was supported by the Deutsche Forschungsgemeinshaft (Grant No. PA 516/7-1). 
\end{acknowledgments}

\bibliography{bib}

\end{document}